\documentclass{article} 

\usepackage{lineno}
\modulolinenumbers[5]
  
\usepackage{appendix}
\usepackage{amsmath,amssymb}
\usepackage[labelfont=bf]{caption}
\usepackage{cases}
\usepackage{array}
\usepackage{graphicx}
\usepackage{dcolumn}
\usepackage{booktabs,hhline}
\usepackage{multirow}
\usepackage{bm}
\usepackage[colorlinks=true,linkcolor=blue,citecolor=blue,urlcolor=blue]{hyperref}
\usepackage[nameinlink]{cleveref}
\usepackage{braket}
\usepackage{soul}
\usepackage{lscape}
\usepackage{orcidlink}
\usepackage{newtxtext}

\usepackage[
backend=biber,
backref=false,
backrefstyle=two,
style=numeric-comp,
sorting=none
]{biblatex}

\DeclareFieldFormat{pages}{#1} 
\DeclareFieldFormat[article,periodical]{volume}{\mkbibbold{#1}} 
\renewbibmacro{in:}{}
\numberwithin{equation}{section}

\newcolumntype{P}[1]{>{\centering\arraybackslash}p{#1}}
 
\usepackage{xcolor}     

\newcommand{\orcid}[1]{\href{https://orcid.org/#1}{\textcolor[HTML]{A6CE39}{\aiOrcid}}}

\begin{document}

\begin{center}
{\Large \bf 
Radial stability of spherical bosonic stars and critical points
}
\vspace{0.8cm}
\\
{
Nuno M. Santos$^{1,2,\orcidlink{0000-0003-3036-7930}}$, Carolina L. Benone$^{3,\orcidlink{0000-0002-3675-9452}}$, Carlos A. R. Herdeiro$^{1,\orcidlink{0000-0002-9619-2013}}$ 
\\
\vspace{0.3cm}
{\small $^1$Departamento de Matem\'atica da Universidade de Aveiro and }\\
{\small  Centre for Research and Development  in Mathematics and Applications (CIDMA),}\\ 
{\small Campus de Santiago, 3810--193 Aveiro, Portugal}\\
\vspace{0.3cm}
{\small $^2$Departamento de F\'\i sica, Instituto Superior T\'ecnico -- IST, Universidade de Lisboa -- UL,} \\
{\small Avenida Rovisco Pais 1, 1049--001 Lisboa, Portugal}\\
\vspace{0.3cm}
{\small $^3$Campus Universitário de Salinópolis, Universidade Federal do Pará, 68721--000, Salinópolis, Pará, Brazil}
}
\end{center}
\vspace{0.3cm}
\begin{abstract}
We study radial perturbations of spherically symmetric spin-$0$ and spin-$1$ bosonic stars, computing numerically the squared frequency of the fundamental mode. We find that not all critical points -- where the Arnowitt--Deser--Misner mass attains an extremum -- correspond to zero modes. Thus, radial stability does not \textit{always} change at such critical points. The results are in agreement with the so-called critical point method.
\end{abstract}


\tableofcontents

\newpage

\section{Introduction}\label{sec:1}

Self-gravity is key to the understanding of the physics of compact objects, such as white dwarfs, neutron stars, and black holes (BHs). Apart from the latter, which cannot support themselves at all against gravitational collapse, the equilibrium of compact objects relies on a balance between the attractive pull of gravity and a repulsive push of some sort. In the case of white dwarfs and neutron stars, the repulsion is nothing but degeneracy pressure of electrons and neutrons, respectively, caused by the Pauli exclusion principle and the Heisenberg uncertainty principle.

If compact objects are composed of something else rather than ordinary baryonic matter, they are said to be exotic. A number of exotic compact objects has been put forward over the past decades (see~\cite{Cardoso:2019rvt} for a review). Examples include boson stars (BSs)~\cite{Kaup:1968zz,Ruffini:1969qy}, anisotropic stars~\cite{Bowers:1974tgi}, wormholes~\cite{Morris:1988tu}, gravastars~\cite{Mazur:2004fk}, or fuzzballs~\cite{Mathur:2005zp}. BSs in particular have attracted much attention recently (see~\cite{Schunck:2003kk,Liebling:2012fv,Visinelli:2021uve} for reviews). As first conceived by Kaup, they are solitonic solutions of Einstein's gravity minimally coupled to a massive, complex scalar field with harmonic time dependence. There is nothing special about scalar fields, though. Such compact objects may also consist of massive, complex vector (or Proca) fields, in which case they are known as Proca stars (PSs)~\cite{Brito:2015pxa}. Collectively, BSs and PSs are  referred to as \textit{bosonic stars}. Unlike white dwarfs and neutron stars, they are supported against gravitational collapse by the Heisenberg uncertainty principle only. From a theoretical standpoint, bosonic stars are particularly interesting because they appear in well-motivated and self-consistent physical theories and have a formation mechanism, known as gravitational cooling~\cite{Seidel:1993zk,Guzman:2006yc}. From an observational standpoint, they stand out from the plethora of exotic compact objects as simple yet astrophysically viable compact objects, which can serve as a model (or a proxy) for lumps of ultralight dark matter~\cite{Suarez:2013iw,Hui:2016ltb,Freitas:2021cfi}. 

Among the most pressing questions around the astrophysical viability of compact objects is that of stability. Bosonic stars are no exception: as hypothetical compact objects, they should be stable with respect to sufficiently small perturbations (or, in case they are unstable, be sufficiently long-lived). The linear stability of BSs was first addressed by Gleiser~\cite{Gleiser:1988rq} and Jetzer~\cite{Jetzer:1988vr} in the 80s, following Chandrasekhar's seminal work on perfect-fluid stars~\cite{Chandrasekhar:1964zz}. Both realized that, like normal stars, the fundamental set of BSs can be either radially stable or unstable, depending on the field value at the star's center $\psi_0$, and estimated upper bounds for the threshold of the instability. Gleiser and Watkins later solved numerically the linear perturbation equations (also referred to as ``pulsation equations") and found that the instability trigger value matched the BS maximum Arnowitt--Deser--Misner (ADM) mass~\cite{Gleiser:1988ih}.  Soon afterward, Seidel and Suen studied the linear and nonlinear radial stability of mini-BSs using numerical relativity~\cite{Seidel:1990jh} and concluded that: stable mini-BSs oscillate (when perturbed), emitting bosonic radiation and losing mass as they relax to a configuration with lower mass and larger radius; unstable mini-BSs, on the other hand, either collapse to a black hole or migrate to a configuration in the stable branch. More recently, this picture was confirmed and extended, in both the time and frequency domains, to mini-BSs in the first excited mode~\cite{Hawley:2000dt} as well as massive-BSs in the fundamental and first excited modes~\cite{Guzman:2004jw,Kain:2021rmk}, and spherically symmetric PSs~\cite{Sanchis-Gual:2017bhw}. It has been reported in particular that some unstable solutions (also in the case of fundamental mini-BSs) disperse altogether due to having more energy than the corresponding collection of particles. 

Radial perturbations correspond to physical, non-radiative degrees of freedom, describing changes in the mass of the system~\cite{Zerilli:1970wzz}. More precisely, they refer to polar (or even or electric) $\ell=0$ perturbations of the equilibrium (spherically symmetric) background in the context of linear perturbation theory. Conclusions drawn from radial linear stability analysis must be taken with a pinch of salt: a compact object that is stable with respect to sufficiently small radial perturbations might be prone to instabilities grown out of an angular disturbance, for instance. Non-radial linear stability of mini-BSs as well as of massive and solitonic BSs has been studied in~\cite{Kojima:1991np,Yoshida:1994xi,Macedo:2013jja}.

\begin{figure}[ht!]
    \centering
    \includegraphics[scale=0.9]{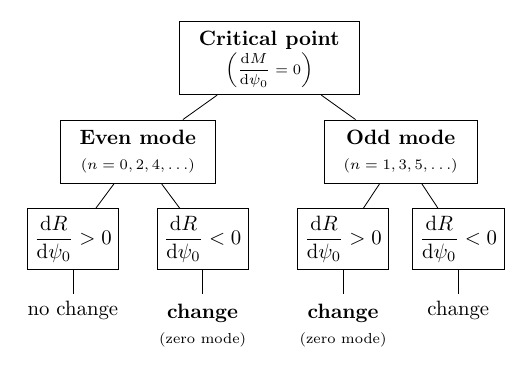}
    \caption{Assessment of stability-instability transitions with the critical point method.}
    \label{fig:1.1}
\end{figure}

It is possible to analyze the radial linear stability of normal stars using the \textit{critical point method}~\cite[\nopp Chap. 6]{Shapiro:1983du}. On a plot of the equilibrium mass $M$ vs. $\psi_0$, changes in stability, dictated by zero-frequency modes, are linked to critical points, i.e. solutions satisfying $\text{d}M/\text{d}\psi_0=0$. The linear perturbation equations pose a Sturm-Liouville boundary value problem on the finite interval $[0,R]$, where $R$ is the star radius, for the perturbation frequency squared $\Omega^2$. According to spectral theory, there are infinitely many real eigenvalues $\Omega_0^2$, $\Omega_1^2$, $\Omega_2^2,\ldots$, and the eigenfunction corresponding to the eigenvalue $\Omega_n^2$, $n\in\mathbb{N}_0$, has exactly $n$ zeros in the open interval $(0,R)$, i.e. $n$ nodes. An eigenfunction with an odd (even) number of nodes is often referred to as an odd (even) mode. Additionally, the eigenvalues are ordered, i.e. $\Omega_0^2<\Omega_1^2<\Omega_2^2<\ldots$. The stability can be analyzed by noting that $\text{d}R/\text{d}\psi_0>0$ ($\text{d}R/\text{d}\psi_0<0$) at a critical point corresponds to a change of sign of an odd (even) mode. The critical point method is summarized schematically in~\autoref{fig:1.1}.

Unlike normal stars, BSs do not have a definite surface, i.e. a boundary outside which the energy density vanishes. In fact, the field extends to spacelike infinity and, for that reason, the corresponding linear perturbation equations form a \textit{singular} Sturm-Liouville boundary value problem. Strictly speaking, the previous result on the eigenvalues cannot be applied to BSs. However, their energy density decay exponentially close to spacelike infinity, where its value is much smaller than the maximum. It is then reasonable to define the ``surface" of a BS as a spherical surface enclosing most of its equilibrium mass. A popular choice is the circumferential radius containing $99\%$ of $M$. Such a definition ensures the validity of the critical point method in this case.

While the critical point method suffices to identify zero-frequency modes and stability-instability transitions across the parameter space, it does not provide any estimates of the perturbation frequencies. These are important to understand the spectra of normal and quasinormal modes of stable BSs. To this end, the most straightforward approach is to (numerically) integrate the perturbation equations. One of the goals of this paper is to present the eigenvalues $\Omega^2_0$ for mini-, solitonic ~\cite{Friedberg:1986tq} and axion~\cite{Guerra:2019srj} BSs as well as for PSs~\cite{Brito:2015pxa}. 

Another purpose of this paper is to clarify some ideas about BS stability present in the literature. An example is the statement~\cite{Siemonsen:2020hcg} that the radial stability of a spherically symmetric BS changes whenever $\text{d}M/\text{d}\psi_0=0$. While it is true that transitions from stable to unstable configurations, and vice versa, only occur at critical points, defined by $\text{d}M/\text{d}\psi_0=0$, the examples herein make clear that not all critical points signal such transitions, as illustrated in~\autoref{fig:1.1}.

The rest of the paper is organized as follows. In~\autoref{sec:2}, the action, equations of motion, and relevant physical quantities of different BS models are introduced. \autoref{sec:3} is devoted to a brief review of the formulation describing spherically symmetric BSs, whereas \autoref{sec:4} formulates their first-order radial perturbations in the Zerilli gauge. In \autoref{sec:5}, the space of solutions of both equilibrium and perturbed solutions is presented in a side-by-side comparison of the different models. An overview of the work is sketched in \autoref{sec:6}, together with some final remarks.

\textit{Notation}. In the following, the speed of light $c$ and the Planck constant $\hbar$ are set equal to one ($c=\hbar=1$), unless stated otherwise. The mostly positive metric signature $(-+++)$ is adopted and Latin letters ($a,b,c,\ldots$) are used for abstract index notation. Since the paper addresses first-order perturbations, every tensor field $A$ can be expanded in powers of a bookkeeping parameter $\epsilon$, $A(\epsilon)=A^{(0)}+\epsilon\,A^{(1)}+\ldots$, where $A^{(0)}$ is the unperturbed value and $A^{(1)}$ is the first-order perturbation of $A(\epsilon)$.
%
\section{Framework}\label{sec:2}
%
Consider the action describing a complex spin-$s$ field, $s\in\{0,1\}$, minimally coupled to Einstein's gravity:
    \begin{equation}
    \label{eq:2.1}
        \mathcal{S}_s = \int\text{d}^4x\,\sqrt{-g}\left[\frac{R}{16\pi G}+\mathcal{L}_s\right]\ ,
	\end{equation}
where $G$ is the gravitational constant, $g_{ab}$ is the metric, with determinant $g$ and Ricci scalar $R$, and the matter Lagrangians read
	\begin{equation}
    \label{eq:2.2}
        \mathcal{L}_0=-g^{ab}\nabla_a\Phi\nabla_b\bar{\Phi}-V_0(\Phi\bar{\Phi})\ .
        \quad
        \mathcal{L}_1=-\frac{1}{4}\mathcal{F}_{ab}\bar{\mathcal{F}}^{ab}-V_1(\mathcal{A}_a\bar{\mathcal{A}}^a)\ ,
	\end{equation}
$\Phi$ is a complex scalar field, $\mathcal{A}_a\text{d}x^a$ is a complex Proca field with strength $\mathcal{F}=\text{d}\mathcal{A}$, and $V_s$ is the corresponding spin-$s$ field potential. Moreover, the overbar $\bar{\cdot}$ denotes complex conjugation.

This paper focuses on three particular forms of the scalar-field potential, namely
    \begin{subnumcases}{V_0(\Phi\bar{\Phi})=}
        \label{eq:2.3a}
        ~\mu^2\Phi\bar{\Phi} & \text{(mini)}\ ,\\
        \label{eq:2.3b}
        ~\mu^2\Phi\bar{\Phi}\left(1-\dfrac{2\Phi\bar{\Phi}}{v_0^2}\right)^2 & \text{(solitonic)}\ ,\\
        \label{eq:2.3c}
        ~\dfrac{2\mu^2f_a^2}{B}\left[1-\sqrt{1-4B\sin^2\left(\dfrac{\sqrt{\Phi\bar{\Phi}}}{2f_a}\right)}\,\right] & \text{(axionic)}\ ,
    \end{subnumcases}
and on the simplest form of the Proca-field potential,
    \begin{equation}
        V_1(\mathcal{A}_a\bar{\mathcal{A}}^a)=\frac{1}{2}\mu^2\mathcal{A}_a\bar{\mathcal{A}}^a\ .
    \end{equation}
Bosonic stars in each model are known as mini-boson stars (MBSs), solitonic boson stars (SBSs), axionic boson stars (ABSs), and Proca stars (PSs), respectively. In the above potentials, $\{\mu,v_0,f_a\}$ are free parameters: $\mu=mc/\hbar$ is the inverse Compton wavelength of the corresponding spin-$s$ field (and $m$ is its bare mass), $v_0$ is the degenerate vacuum of the solitonic potential~\eqref{eq:2.3b} and $f_a$ is the Peccei-Quinn symmetry breaking scale. Additionally, $B=z/(1+z)^2\approx0.22$, with $z\equiv m_u/m_d\approx0.48$ being the up-quark-to-down-quark mass ratio. The second term in Eq.~\eqref{eq:2.3c} is the standard QCD axion potential, which is non-zero when $\Phi=0$. The first term is added so that $V_0(0)=0$ and hence asymptotic flatness is ensured. Note that the potential in Eq.~\eqref{eq:2.3a} is the linear approximation of both Eqs~\eqref{eq:2.3b}--\eqref{eq:2.3c} around\footnote{For the latter case, the linear approximation is valid when $\sqrt{\Phi\bar{\Phi}}\ll f_a$.} $\Phi\bar{\Phi}=0$.

Varying the action in Eq.~\eqref{eq:2.1} with respect to the metric and to the matter field yields the equations of motion, namely Einstein equations, 
    \begin{equation}
    \label{eq:2.5}
        E_{ab}\equiv G_{ab}-8\pi GT_{ab}^{[s]}=0\ ,
    \end{equation}
where $G_{ab}$ is the Einstein tensor, and $T_{ab}^{[s]}$ is the energy-momentum tensor of the spin-$s$ field,
	\begin{subequations}
    \begin{align}
        &T_{ab}^{[0]}=\nabla_a\Phi\nabla_b\bar{\Phi}+\nabla_b\Phi\nabla_a\bar{\Phi}-g_{ab}\left[\frac{1}{2}g^{cd}\left(\nabla_c\Phi\nabla_d\bar{\Phi}+\nabla_d\Phi\nabla_c\bar{\Phi}\right)+V_0(\Phi\bar{\Phi})\right]\ ,\\
        &T_{ab}^{[1]}=\frac{1}{2}g^{cd}\left(\mathcal{F}_{ac}\bar{\mathcal{F}}_{bd}+\bar{\mathcal{F}}_{ac}\mathcal{F}_{bd}\right)-\frac{1}{4}g_{ab}F_{cd}\bar{F}^{cd}+\frac{1}{2}\mu^2\left(\mathcal{A}_a\bar{\mathcal{A}}_b+\bar{\mathcal{A}}_a \mathcal{A}_b-g_{ab}\mathcal{A}_c\bar{\mathcal{A}}^c\right)\ ,
    \end{align}
	\end{subequations}
and the matter equations
    \begin{subequations}
    \label{eq:2.6}
    \begin{align}
        \label{eq:2.6a}
        &\nabla_a\nabla^a\Phi-\frac{\partial V_0}{\partial(\Phi\bar{\Phi})}\,\Phi=0\ ,\\
        \label{eq:2.6b}
        &\nabla_a \mathcal{F}^{ab}-\mu^2\mathcal{A}^b=0\ .
    \end{align}
    \end{subequations}
Equation~\eqref{eq:2.6a} is the Klein-Gordon equation, whereas Eq.~\eqref{eq:2.6b} is the Proca equation. The latter implies the Lorenz condition $\nabla_a \mathcal{A}^a=0$.

In either case, the action in Eq.~\eqref{eq:2.1} possesses a global $U(1)$ symmetry, i.e. it is invariant under the transformation $\{\Phi,\mathcal{A}_a\}\rightarrow e^{i\chi}\{\Phi,\mathcal{A}_a\}$, with $\chi$ constant. This implies the existence of the four-currents
	\begin{align}
    \label{eq:2.4}
        j^a_0=i(\Phi\nabla^a\bar{\Phi}-\bar{\Phi}\nabla^a\Phi)\ ,
        \qquad
        j^a_1=\frac{i}{2}\left(\bar{\mathcal{F}}^{ab}\mathcal{A}_b-\mathcal{F}^{ab}\bar{\mathcal{A}}_b\right)\ ,
	\end{align}
which are conserved, i.e. $\nabla_a{j_s^a}=0$. There exists a Noether charge $Q_s$, obtained by integrating the timelike component of the four-currents on a spacelike surface $\Sigma$,
    \begin{align}
        Q_s=\int_\Sigma\text{d}^3x~j_s^0\ .
	\end{align}
Upon quantization, $Q$ is nothing but the particle number.

    The Komar mass reads
    \begin{align}
    \label{eq:2.9}
        M_s=\frac{1}{4\pi G}\int_\Sigma \text{d}V\,R_{ab}n^a\xi^b=2\int_\Sigma\text{d}V\left(T_{ab}^{[s]}-\frac{1}{2}g_{ab}T^{[s]}\right)n^a\xi^b\ ,
    \end{align}
where $\Sigma$ is an asymptotically-flat spacelike hypersurface, $n^\alpha$ is a future-pointing unit vector normal to $\Sigma$, $\text{d}V$ is the $3$-volume form induced on $\Sigma$, and $T^{[s]}=g^{ab}T_{ab}^{[s]}$ is the trace of the energy-momentum tensor $T_{ab}^{[s]}$.

\section{Equilibrium solutions}\label{sec:3}

In this paper, only radial perturbations of spherically symmetric bosonic stars will be considered. For completeness, the construction and physical properties of the equilibrium solutions are reviewed in the following. The solutions are parametrized by the spin-$s$ field value at the star's center $\psi_0$. This section follows closely the definitions and conventions in~\cite{Herdeiro:2017fhv}. 

\subsection{Ans\"{a}tze and equations of motion}\label{sec:3.1}
%
A line element compatible with a static spherically-symmetric spacetime is
	\begin{equation}  
    \label{eq:3.1}
        \text{d}s^2=g_{ab}^{(0)}\,\text{d}x^a\text{d}x^b=-\sigma(r)^2N(r)\text{d}t^2+\frac{\text{d}r^2}{N(r)}+r^2\left( \text{d}\theta^2+\sin^2\theta\,\text{d}\varphi^2\right)\ ,
        \quad
        N(r)=1-\frac{2G\mathcal{M}(r)}{r}\ ,
	\end{equation}
where $\mathcal{M}$ is the Misner--Sharp mass function. Unlike the metric tensor, the matter fields are assumed to have a harmonic time-dependence:
    \begin{subequations}
    \label{eq:3.2}
    \begin{align}
        &\Phi^{(0)}=e^{-i\omega t}\,\phi(r)\ ,\\
        &\mathcal{A}^{(0)}_a\text{d}x^a=e^{-i\omega t}\big[f(r)\,\text{d}t+ig(r)\,\text{d}r\big]\ ,
	\end{align}	 
    \end{subequations}
where $\{\phi,f,g\}$ are real functions. Without loss of generality, one assumes $\omega\in\mathbb{R}^+$.


Restricted to the above ans\"{a}tze, the equations of motion in Eqs.~\eqref{eq:2.5} and~\eqref{eq:2.6} yield a system of three (four) coupled ordinary differential equations for the scalar (Proca) field. The only non-zero components of $E_{ab}$ are $E_{tt}$, $E_{rr}$, $E_{\theta\theta}$ and $E_{\varphi\varphi}$. The equations governing the metric functions $N$ and $\sigma$ read
    \begin{subequations}
    \label{eq:3.3}
    \begin{align}
        \label{eq:3.3a}
        &\frac{1}{r^2\sigma}\partial_r(r\sigma N)-\frac{1}{r^2}+8\pi G\mathcal{V}_s=0\ ,\\
        \label{eq:3.3b}
        &\frac{\partial_r\sigma}{\sigma}-8\pi G\mathcal{W}_s=0\ ,
    \end{align}    
    \end{subequations}
where
    \begin{equation}
        \mathcal{V}_0=V_0 ,
        \qquad
        \mathcal{V}_1=\frac{1}{2}\left(\frac{\partial_r f-\omega g}{\sigma}\right)^2\ ,
    \end{equation}
and
    \begin{equation}
        \mathcal{W}_0=\left[(\partial_r\phi)^2+\frac{\omega^2\phi^2}{\sigma^2N^2}\right]r\ ,
        \qquad
        \mathcal{W}_1=\frac{\mu^2}{2}\left(\frac{f^2}{\sigma^2N}+g^2\right)r\ .
    \end{equation}

On the other hand, the matter equations become
    \begin{subequations}
    \label{eq:3.6}
    \begin{align}
        &\frac{1}{r^2\sigma}\partial_r(r^2\sigma N\partial_r\phi)+\left(\frac{\omega^2}{\sigma^2N}-V_0'\right)\phi=0\ ,\\
        &
        \partial_r\left[\frac{r^2}{\sigma}(\partial_rf-\omega g)\right]-\mu^2\frac{r^2f}{\sigma N}=0\ ,
        \quad
        \omega(\partial_rf-\omega g)+\mu^2\sigma^2Ng=0\ ,
    \end{align}    
    \end{subequations}
where $V_0^{(k)}\equiv\text{d}^kV_0(\phi^2)/\text{d}(\phi^2)^k$, $k\in\mathbb{N}$. The Lorenz condition reads
    \begin{equation}
        \frac{1}{r^2}\partial_r\left(r^2\sigma Ng\right)+\frac{\omega f}{\sigma N}=0\ .
    \end{equation}	 

\subsection{Boundary conditions\label{sec:3.2}}
%
Equations~\eqref{eq:3.3} together with Eqs.~\eqref{eq:3.6} can be (numerically) integrated under the boundary conditions (BCs) summarized in~\autoref{tab:1}. The inner BCs result from the smoothness at $r=0$, required to avoid the poles there. In fact, it can be shown that 
    \begin{subequations}
    \label{eq:3.4}
    \begin{align}
        \label{eq:3.4a}
        &\mathcal{M}(r)=\frac{4\pi G}{3}\left[\omega^2\frac{\phi_0^2}{\sigma_0^2}+\Tilde{V}_0\right]r^2+\ldots\ ,\\
        \label{eq:3.4b}
        &\sigma(r)=\sigma_0+\frac{4\pi G\phi_0^2}{\sigma_0}\omega^2r^2+\ldots\ ,\\
        \label{eq:3.4c}
        &\phi(r)=\phi_0-\frac{1}{6}\left[\omega^2\frac{\phi_0}{\sigma_0^2}-\Tilde{V}_0\right]r^2+\ldots\ ,\\
        &f(r)=f_0\left[1+\frac{1}{6}\left(\mu^2-\frac{\omega^2}{\sigma_0^2}\right)\right]r^2+\ldots\ ,\\
        &g(r)=-\frac{\omega}{3}\frac{f_0}{\sigma_0^2}r+\ldots\ ,
    \end{align}
    \end{subequations}
where $\sigma_0\equiv\sigma(0)$, $\phi_0\equiv\phi(0)$ and $f_0\equiv f(0)$ are as yet undetermined constants, and $\Tilde{V}_0\equiv V_0(\phi_0^2)$. Without loss of generality, one can assume that $\phi_0,\,f_0\in\mathbb{R}^+$ thanks to the $\mathbb{Z}_2$-symmetry of the spin-$s$ fields. Moreover, note that $m'(0)=\sigma'(0)=0$.

On the other hand, the outer BCs ($r=\infty$) follow on from the requirement of asymptotic flatness. The leading-order asymptotic behavior of the functions is  
    \begin{subequations}
    \label{eq:3.5}
    \begin{align}
        \label{eq:3.5a}
        & N(r)=1-\frac{2GM}{r}+\ldots\ ,\\
        \label{eq:3.5b}
		&\log\sigma(r)=-\frac{c_0^2}{2}\frac{\mu^2\omega^2}{(\mu^2-\omega^2)^{3/2}}\frac{e^{-2r\sqrt{\mu^2-\omega^2}}}{r}+\ldots\ ,\\
		\label{eq:3.5c}
        &\phi(r)=\phi_\infty\frac{e^{-r\sqrt{\mu^2-\omega^2}}}{r}+\ldots\ ,\\
        &f(r)=f_\infty\frac{e^{-r\sqrt{\mu^2-\omega^2}}}{r}+\ldots\ ,\\
        &g(r)=f_\infty\frac{\omega}{\sqrt{\mu^2-\omega^2}}\frac{e^{-r\sqrt{\mu^2-\omega^2}}}{r}+\ldots\ ,
	\end{align}
    \end{subequations}
where $\{c_0,\phi_\infty,f_\infty\}$ are constants.

\begin{table}[ht!]
    \begin{center}
    \begin{tabular}{lP{1.1cm}P{1.1cm}||P{1.1cm}P{1.1cm}|P{1.1cm}P{1.1cm}P{1.1cm}}
        \cmidrule[\heavyrulewidth]{2-8}
        & \multicolumn{2}{c||}{Metric functions} & \multicolumn{5}{c}{Matter functions}\\
        \cmidrule[\heavyrulewidth]{2-8}
        & $\mathcal{M}$ & $\sigma$ & $\phi$ & $\phi'$ & $f$ & $f'$ & $g$\\
        \midrule
        \multicolumn{1}{l}{Inner BCs ($r=0$)} & 0 & $\sigma_0$ & $\phi_0$ & 0 & $f_0$ & 0 & 0\\
        \multicolumn{1}{l}{Outer BCs ($r=\infty$)} & $M$ & $1$ & 0 & 0 & 0 & 0 & 0\\
        \midrule[\heavyrulewidth]
        & $H_0$ & $H_2$ & $\phi_\pm$ & $\phi'_\pm$ & $f_\pm$ & $f'_\pm$ & $g_\pm$\\
        \midrule
        \multicolumn{1}{l}{Inner BCs ($r=0$)} & $h_0$ & $0$ & $\phi_\pm(0)$ & 0 & $f_\pm(0)$ & 0 & 0\\
        \multicolumn{1}{l}{Outer BCs ($r=\infty$)} & $h_\infty$ & $0$ & 0 & 0 & 0 & 0 & 0\\
        \bottomrule
    \end{tabular}
    \end{center}
    \caption{Boundary conditions (BCs) for both metric and matter functions of equilibrium and perturbed bosonic stars (see~\autoref{sec:3.1} and~\autoref{sec:4.2} for definitions).}
    \label{tab:1}
\end{table}

%
\subsection{Physical quantities}\label{sec:3.3}
%
    The Noether charge~\eqref{eq:2.5} of the equilibrium solutions read
    \begin{subequations}
        \begin{align}
            &Q_0^{(0)}=8\pi\int_0^{\infty}\text{d}r~r^2\frac{\omega \phi^2}{\sigma N}\ ,\\
            &Q_1^{(0)}=4\pi\int_0^{\infty}\text{d}r~r^2\frac{g}{\sigma}(\partial_r f-\omega g)\ ,
	\end{align}
    \end{subequations}
whereas the Komar mass~\eqref{eq:2.6} yield
    \begin{subequations}
        \begin{align}
            &M_0^{(0)}=4\pi\int_0^{\infty}\text{d}r~\frac{r^2}{\sigma N}\left[4\left(\omega^2-\frac{\mu^2}{2}\sigma^2N\right)\phi^2\right]\ ,\\
            &M_1^{(0)}=4\pi\int_0^{\infty}\text{d}r~\frac{r^2}{\sigma N}\left[N(\partial_r f-\omega g)^2+2\mu^2 f^2\right]\ .
	\end{align}
    \end{subequations}
Moreover, it is useful to introduce a definition for the radius. Unlike normal stars, such quantity is ill-defined due to the Yukawa-like asymptotic behavior of the matter fields. Here, the most common definition is adopted: the bosonic stars radius $R$ is the circumferential radius containing $99\%$ of the ADM mass.\footnote{Different definitions can be found in the literature~\cite[\nopp II.C]{Schunck:2003kk}.} The compactness of the equilibrium bosonic stars is then $\mathcal{C}\equiv GM/(c^2R)$.

\section{Perturbed solutions}\label{sec:4}
%
In a spherically symmetric spacetime, metric perturbations can be written as a sum of 10 tensor spherical harmonics of degree $\ell\in\mathbb{N}_0$ and order $|\Tilde{m}|\leq\ell$, which form an orthonormal basis. They can in particular be separated into polar (or even or electric) and axial (or odd or magnetic) perturbations according to their properties under parity transformations (see, e.g.,~\cite[\nopp Chap. 12]{Maggiore:2018sht}). This section addresses radial perturbations of equilibrium bosonic stars. This amounts to considering polar $\ell=0$ perturbations (also known as monopolar perturbations). Although in vacuum $\ell=0$ perturbations can be removed by a gauge transformation (and do not contribute to radiative degrees of freedom of the gravitational field), they are not spurious in this case.  
%
\subsection{Ans\"{a}tze and equations of motion}\label{sec:4.1}
    
    In the Zerilli gauge~\cite{Zerilli:1970wzz}, the metric perturbation reads
    \begin{align}
        \label{eq:4.1}
        g_{ab}^{(1)}\text{d}x^a\text{d}x^b=\sigma(r)^2N(r)\Tilde{H}_0(t,r)\,\text{d}t^2+\frac{\Tilde{H}_2(t,r)}{N(r)}\,\text{d}r^2\ ,
    \end{align}
while matter perturbations are assumed to have the form
    \begin{subequations}
    \label{eq:4.2}
    \begin{align}
        &\Phi^{(1)}=e^{-i\omega t}\,\phi_1(t,r)\ ,\\
        &\mathcal{A}_a^{(1)}\text{d}x^a=e^{-i\omega t}\left[f_1(t,r)\,\text{d}t+ig_1(t,r)\,\text{d}r\right]\ ,
	\end{align}	 
    \end{subequations}
%
%
where $\{\Tilde{H}_0,\Tilde{H}_2\}$ are real functions, whereas $\{\phi_1,f_1,g_1\}$ are complex functions.

The first-order equations of motion include terms that prevent the existence of monochromatic (single-frequency) solutions. The simplest solutions are a superposition of two monochromatic waves with frequencies $\omega_\pm=\omega\pm\Omega$, where $\Omega$ is the perturbation frequency. The perturbation functions are of the form
    \begin{subequations}
    \begin{align}
        \Tilde{H}_0(t,r)&=(e^{-i\Omega t}+e^{+i\Omega t})H_0(r)\ ,\\
        \Tilde{H}_2(t,r)&=(e^{-i\Omega t}+e^{+i\Omega t})H_2(r)\ ,\\
        \phi_1(t,r)&=e^{-i\Omega t}\,\phi_+(r)+e^{+i\Omega t}\,\phi_-(r)\ ,\\
        f_1(t,r)&=e^{-i\Omega t}\,f_+(r)+e^{+i\Omega t}\,f_-(r)\ ,\\
        g_1(t,r)&=e^{-i\Omega t}\,g_+(r)+e^{+i\Omega t}\,g_-(r)\ .
    \end{align}
    \end{subequations}
where $\{H_0,\,H_2,\,\phi_\pm,\,f_\pm,\,g_\pm\}$ are real functions. The resulting first-order equations of motion become an eigenvalue problem for $\Omega^2$, which is real, i.e. $\Omega$ is either purely real or purely imaginary, the perturbed solution being either stable or unstable, respectively. Plugging Eqs.~\eqref{eq:4.1} and~\eqref{eq:4.2} into ${E^t}_t$, ${E^t}_r$ and ${E^r}_r$, and keeping linear terms in the bookkeeping parameter $\epsilon$, one gets, respectively,
    \begin{subequations}
    \label{eq:4.4}
    \begin{align}
        \label{eq:4.4a}
        &\frac{1}{r}\partial_r(NH_2)+\frac{N}{r^2}H_2-8\pi G\mathcal{X}_s=0\ ,\\
        \label{eq:4.4b}
        &\Omega H_2-8\pi G r\mathcal{Y}_s=0\ ,\\
        \label{eq:4.4c}
        &\frac{N}{r}\partial_rH_0+\frac{N}{r}\left(\frac{1}{r}+\frac{\partial_rN}{N}+\frac{2\partial_r\sigma}{\sigma}\right)H_2+8\pi G\mathcal{Z}_s=0\ ,
    \end{align}
    \end{subequations}
where
    \begin{subequations}
    \begin{align}
        &\mathcal{X}_0=\frac{\omega^2\phi^2}{\sigma^2N}H_0-N(\partial_r\phi)^2H_2+N(\partial_r\phi)\partial_r(\phi_++\phi_-)+\phi\left(\frac{\omega\omega_+}{\sigma^2N}+V_0'\right)\phi_++\phi\left(\frac{\omega\omega_-}{\sigma^2N}+V_0'\right)\phi_-\ ,\\
        &\mathcal{Y}_0=(\partial_r\phi)(\omega_+\phi_+-\omega_-\phi_-)-\omega\phi\,\partial_r(\phi_+-\phi_-)\ ,\\
        &\mathcal{Z}_{0}=\mathcal{X}_0-2\phi V_0'(\phi_++\phi_-)\ ,\\
        &\mathcal{X}_1=\frac{\mu^2}{2}\frac{f}{\sigma^2N}(fH_0+f_++f_-)-\frac{\mu^2}{2}Ng^2H_2-\frac{1}{2}\left[\frac{\omega_+(\partial_rf-\omega g)}{\sigma^2}-\mu^2Ng\right]g_+-\nonumber\\
        &\phantom{\mathcal{X}_1=\frac{\mu^2}{2}\frac{f}{\sigma^2N}}
        -\frac{1}{2}\left[\frac{\omega_-(\partial_rf-\omega g)}{\sigma^2}-\mu^2Ng\right]g_-+\frac{\partial_r f-\omega g}{2\sigma^2}\left[(\partial_r f-\omega g)(H_0-H_2)+\partial_r(f_++f_-)\right]\ ,\\
        &\mathcal{Y}_1=\frac{\mu^2}{2}\left[g(f_+-f_-)-f(g_+-g_-)\right]\ ,\\
        &\mathcal{Z}_1=\mu^2Ng(g_++g_-)-\mathcal{X}_1
    \end{align}
    \end{subequations}
When $\mathcal{X}_s=\mathcal{Y}_s=\mathcal{Z}_s=0$, Eq.~\eqref{eq:4.4b} dictates that $\Omega=0$, and one obtains from Eq.~\eqref{eq:4.4a} that $H_2(r)\propto(rN)^{-1}$. On the other hand, close to the outer boundary, Eq.~\eqref{eq:4.4c} retrieves $H_0(r)=a+H_2(r)$, where $a\in\mathbb{R}$ is a constant. 
The first-order Klein-Gordon equation reads
    \begin{align}
        \label{eq:4.6}
        &\frac{1}{r^2\sigma}\partial_r(r^2\sigma N\partial_r\phi_\pm)+\left(\frac{\omega_\pm^2}{\sigma^2N}-V_0'-\phi^2V_0''\right)\phi_\pm=\nonumber\\
        &\qquad=\phi^2V_0''\phi_\mp+\frac{1}{2}N\partial_r\phi\,\partial_r(H_0+H_2)-\frac{\omega\phi}{2\sigma^2N}\left[(\omega+\omega_\pm)H_0+(\omega-\omega_\mp) H_2\right]+\frac{H_2}{r^2\sigma}\partial_r(r^2\sigma N\partial_r\phi)\nonumber\\
        &\qquad=\phi^2V_0''\phi_\mp+\frac{1}{2}N\partial_r\phi\,\partial_r(H_0+H_2)-\frac{\omega(\omega+\omega_\pm)\phi}{2\sigma^2N}(H_0+H_2)+\phi V_0'H_2\ ,
    \end{align}
whereas the first-order Proca equation yields
    \begin{subequations}
    \label{eq:4.7}
    \begin{align}
        &\frac{1}{r^2\sigma}\partial_r\left[\frac{r^2}{\sigma}(\partial_rf_\pm-\omega_\pm g_\pm)\right]+\frac{H_0-H_2}{r^2\sigma}\partial_r\left[\frac{r^2}{\sigma}(\partial_r f-\omega g)\right]+\frac{\partial_r f-\omega g}{2\sigma^2}\partial_r(H_0-H_2)-\frac{\mu^2}{\sigma^2N}(fH_0+f_\pm)=\ ,\nonumber\\
        &\qquad=\frac{1}{r^2\sigma}\partial_r\left[\frac{r^2}{\sigma}(\partial_rf_\pm-\omega_\pm g_\pm)\right]+\frac{\partial_r f-\omega g}{2\sigma^2}\partial_r(H_0-H_2)-\frac{\mu^2}{\sigma^2N}(fH_2+f_\pm)=0\ ,\\
        &\omega_\pm(\partial_rf_\pm-\omega_\pm g_\pm)+\frac{\omega+\omega_\pm}{2}(\partial_r f-\omega g)(H_0-H_2)+\mu^2\sigma^2N(g_\pm-g H_2)\nonumber\\
        &\qquad=\omega_\pm(\partial_rf_\pm-\omega_\pm g_\pm)+\mu^2\sigma^2N(g_\pm-g H_0)\mp\frac{1}{2}\frac{\omega_+-\omega_-}{\omega_++\omega_-}\mu^2\sigma^2Ng(H_0-H_2)=0\ ,
    \end{align}
    \end{subequations}
Finally, the first-order Lorenz condition is
    \begin{subequations}
    \begin{align}
        &\frac{1}{r^2}\partial_r(r^2\sigma Ng_\pm)-\frac{H_2}{r^2}\partial_r(r^2\sigma Ng)+\frac{\omega_\pm f_\pm}{\sigma N}+\frac{(\omega+\omega_\pm)f}{2\sigma N}H_0-\frac{(\omega-\omega_\pm)f}{2\sigma N}H_2-\frac{1}{2}Ng\partial_r(H_0+H_2)=\nonumber\\
        &\qquad=\frac{1}{r^2}\partial_r(r^2\sigma Ng_\pm)+\frac{\omega_\pm f_\pm}{\sigma N}+\frac{(\omega+\omega_\pm)f}{2\sigma N}(H_0+H_2)-\frac{1}{2}Ng\partial_r(H_0+H_2)=0\ .
    \end{align}
    \end{subequations}

\subsection{Boundary conditions}\label{sec:4.2}

The perturbed functions $\{H_0,\,H_2,\,\phi_\pm,\,f_\pm,\,g_\pm\}$ must have regular Taylor series around $r=0$. Thus, the inner boundary conditions are 
\begin{subequations}
    \begin{align*}
        &H_0(r)=
        \left\{
        \begin{array}{ll}
           h_0-\dfrac{8\pi G}{3}\phi_0\left\{\dfrac{2\omega^2\phi_0}{\sigma_0^2}h_0+\left[\dfrac{2\omega\omega_+}{\sigma_0^2}-\Tilde{V}_0'\right]\phi_+(0)+\left[\dfrac{2\omega\omega_-}{\sigma_0^2}-\Tilde{V}_0'\right]\phi_-(0)\right\}r^2+\ldots\ , & \quad s=0\ , \\ \\
           h_0-\dfrac{8\pi G}{3}\dfrac{\mu^2f_0}{\sigma_0^2}[f_0h_0+f_+(0)+f_-(0)]r^2+\ldots\ , & \quad s=1\ ,
        \end{array}
        \right. \\
        &H_2(r)=
        \left\{
        \begin{array}{ll}
           \dfrac{8\pi G}{3}\phi_0\left\{\dfrac{\omega^2\phi_0}{\sigma_0^2}h_0+\left[\dfrac{\omega\omega_+}{\sigma_0^2}+\Tilde{V}_0'\right]\phi_+(0)+\left[\dfrac{\omega\omega_-}{\sigma_0^2}+\Tilde{V}_0'\right]\phi_-(0)\right\}r^2+\ldots\ , & \quad s=0\ , \\ \\
           \dfrac{4\pi G}{3}\dfrac{\mu^2f_0}{\sigma_0^2}\left[f_0h_0+f_+(0)+f_-(0)\right]r^2+\ldots\ , & \quad s=1 \ ,
        \end{array}
        \right. \\
        &\phi_\pm(r)=\phi_\pm(0)-\frac{1}{3}\left\{\frac{\omega(\omega+\omega_\pm)\phi_0}{4\sigma_0^2}h_0+\frac{1}{2}\left[\frac{\omega_\pm^2}{\sigma_0^2}-\Tilde{V}_0'-\phi_0^2\Tilde{V}_0''\right]\phi_\pm(0)-\frac{1}{2}\phi_0^2\Tilde{V}_0''\phi_\mp(0)\right\}r^2+\ldots\ ,\\
        &f_\pm(r)=f_\pm(0)-\frac{1}{6}\left[\left(\frac{\omega_\pm^2}{\sigma_0^2}-\mu^2\right)f_\pm(0)+\frac{\omega_\pm(\omega+\omega_\pm)}{2}f_0h_0\right]r^2+\ldots\ ,\\
        &g_\pm(r)=-\frac{1}{3\sigma_0^2}\left[\omega_\pm f_\pm(0)+\frac{\omega+\omega_\pm}{2}f_0h_0\right]r+\ldots\ ,
        \end{align*}
    \end{subequations}
where $\{h_0,\,\phi_\pm(0),\,f_\pm(0)\}$ are as yet undetermined constants, $\Tilde{V}_0'\equiv V'(\phi_0^2)$ and $\Tilde{V}_0''\equiv V''(\phi_0^2)$. The first-order matter equations are linear, which means that both $\phi_+(0)$ and $f_+(0)$ (say) can be set to unity without loss of generality. Since $\phi_+=\phi_-$ and $f_+=f_-$ when $\Omega=0$, one sets $\phi_+(0)=\phi_-(0)$ and $f_+(0)=f_-(0)$ and checks afterwards whether the outer boundary conditions hold,
    \begin{subequations}
    \begin{align}
        &H_0(r)\approx h_\infty+\frac{a_\infty}{r}\ ,\\
        &H_2(r)\approx \frac{a_\infty}{r}\ ,\\
        &\phi_\pm(r)\approx a_\pm\,\frac{e^{-r\sqrt{\mu^2-\omega_\pm^2}}}{r}\ ,\\
        &f_\pm(r)\approx b_\pm\,\frac{e^{-r\sqrt{\mu^2-\omega_\pm^2}}}{r}\ ,\\
        &g_\pm(r)\approx b_\pm\,\frac{\omega_\pm}{\sqrt{\mu^2-\omega_\pm^2}}\frac{e^{-r\sqrt{\mu^2-\omega_\pm^2}}}{r}\ .
    \end{align}
    \end{subequations}
where $\{h_\infty,a_\infty,a_\pm,b_\pm\}$ are constants. The BCs are summarized in~\autoref{tab:1}.
%

\subsection{Physical quantities}\label{sec:4.3}

The Noether charge~\eqref{eq:2.5} and Komar mass~\eqref{eq:2.6} of the perturbed solutions are time-dependent, yielding, respectively, $Q_s=Q_s^{(0)}+\epsilon\,Q_s^{(1)}+\ldots$ and $M_s=M_s^{(0)}+\epsilon\,M_s^{(1)}+\ldots$, where
    \begin{subequations}
    \begin{align}
        &Q_s^{(1)}=4\pi\int_0^\infty\text{d}r\,r^2\rho_{Q_s}^{(1)}\ ,\\
        &M_s^{(1)}=4\pi\int_0^\infty\text{d}r\,r^2\rho_{M_s}^{(1)}=8\pi\cos(\Omega t)\int_0^\infty\text{d}r\,r^2\mathcal{X}_s\ ,
    \end{align}
    \end{subequations}
and
    \begin{subequations}
    \begin{align}
        &\rho_{Q_0}^{(1)}=2\cos(\Omega t)\frac{\phi}{\sigma^2N}\left[2\omega\phi H_0+(\omega+\omega_+)\phi_++(\omega+\omega_-)\phi_-\right]\ ,\\
        &\rho_{Q_1}^{(1)}=-\frac{\cos(\Omega t)}{\sigma^2}\left\{g\left[(\partial_rf_+-\omega_+g_+)+(\partial_rf_--\omega_-g_-)\right]+(\partial_r f-\omega g)[g_++g_-+2g(H_0-H_2)]\right\}\ ,
    \end{align} 
    \end{subequations}

The radial perturbations should leave the Noether charge unchanged, i.e. $Q_s=Q_s^{(0)}$, and thus $Q_s^{(1)}=0$.

\section{Results}\label{sec:5}
%
The numerical results are obtained using \textsc{Mathematica}~\cite{Mathematica}, namely the built-in symbols \textsc{NDSolve} and \textsc{FindRoot}. The results are presented in terms of dimensionless quantities, obtained from products or quotients of the mass $m$ of the bosonic field. They are thus valid for any value of $m$. Assigning a specific value to $m$ sets the characteristic scale of the system. To compare the physical properties of bosonic stars to those of other compact objects, one can express $\{M,Q,R,\omega,\Omega^2\}$ in convenient units (where $\hbar$ and $c$ are reinstated for clarity):
    \begin{subequations}
    \begin{align}
        & M\approx 1.336\left(\frac{10^{-10}~\text{eV}/c^2}{m}\right)\times\frac{M}{m_\text{P}^2/m}\quad[M_\odot]\ ,\\
        & Q\approx2.475\left(\frac{10^{16}~\text{eV}/c^2}{m}\right)^2\times\frac{Q}{m_\text{P}^2/m^2}\quad[\text{mol}]\ ,\\
        &R\approx1.973\left(\frac{10^{-10}~\text{eV}/c^2}{m}\right)\times \frac{R}{\hbar/(mc)}\quad[\text{km}]\ ,\\
        &\omega\approx 151.9\left(\frac{m}{10^{-10}~\text{eV}/c^2}\right)\times\frac{\omega}{mc^2/\hbar}\quad[\text{kHz}]\ ,\\
        &\Omega^2\approx 23082\left(\frac{m}{10^{-10}~\text{eV}/c^2}\right)^2\times\left(\frac{\Omega}{mc^2/\hbar}\right)^2\quad[\text{kHz}^2]\ ,
    \end{align}
    \end{subequations}
where $m_\text{P}=\sqrt{\hbar c/G}$ is the Planck mass. The values of the (dimensionless) quantities to the right of the multiplication signs can be readily read from the plots for a given solution. If $M/(m_\text{P}^2/m)\sim\mathcal{O}(1)$ (say), bosonic fields with $m\lesssim 10^{-10}$ eV$/c^2$ are compatible with stellar-mass or even supermassive objects. 

\subsection{Equilibrium solutions}\label{sec:5.1}
%
To find equilibrium bosonic stars, one fixes $\omega<\mu$, provides an initial guess for $\psi_0$ (either $\phi_0$ or $f_0$), and solves the boundary value problem with the corresponding BCs. This amounts to numerically integrating Eqs.~\eqref{eq:3.3} and \eqref{eq:3.6} from $r\mu=\delta\ll 1$ (here, $\delta=10^{-6}$) to $D\mu\sim\mathcal{O}(10)$, typically, where $D$ is the (numerical) radial coordinate of the outer boundary.\footnote{The numerical value of $D$ should be such that the hypersurface $r=D$ encloses more than $99\%$ of the mass, i.e. $D>R$. The stars become dilute when approaching the Newtonian limit ($\omega\rightarrow\mu$), in which one takes $D\mu\sim\mathcal{O}(100)$.} Once a solution is found, the mass, Noether charge, radius, and compactness can be readily computed. The solutions presented herein are uniquely identified by the field's central value, thus being particularly convenient to display most physical quantities as functions of either $\phi_0$ or $f_0$. 

\begin{figure}[ht!]
    \centering
    \includegraphics[scale=0.28]{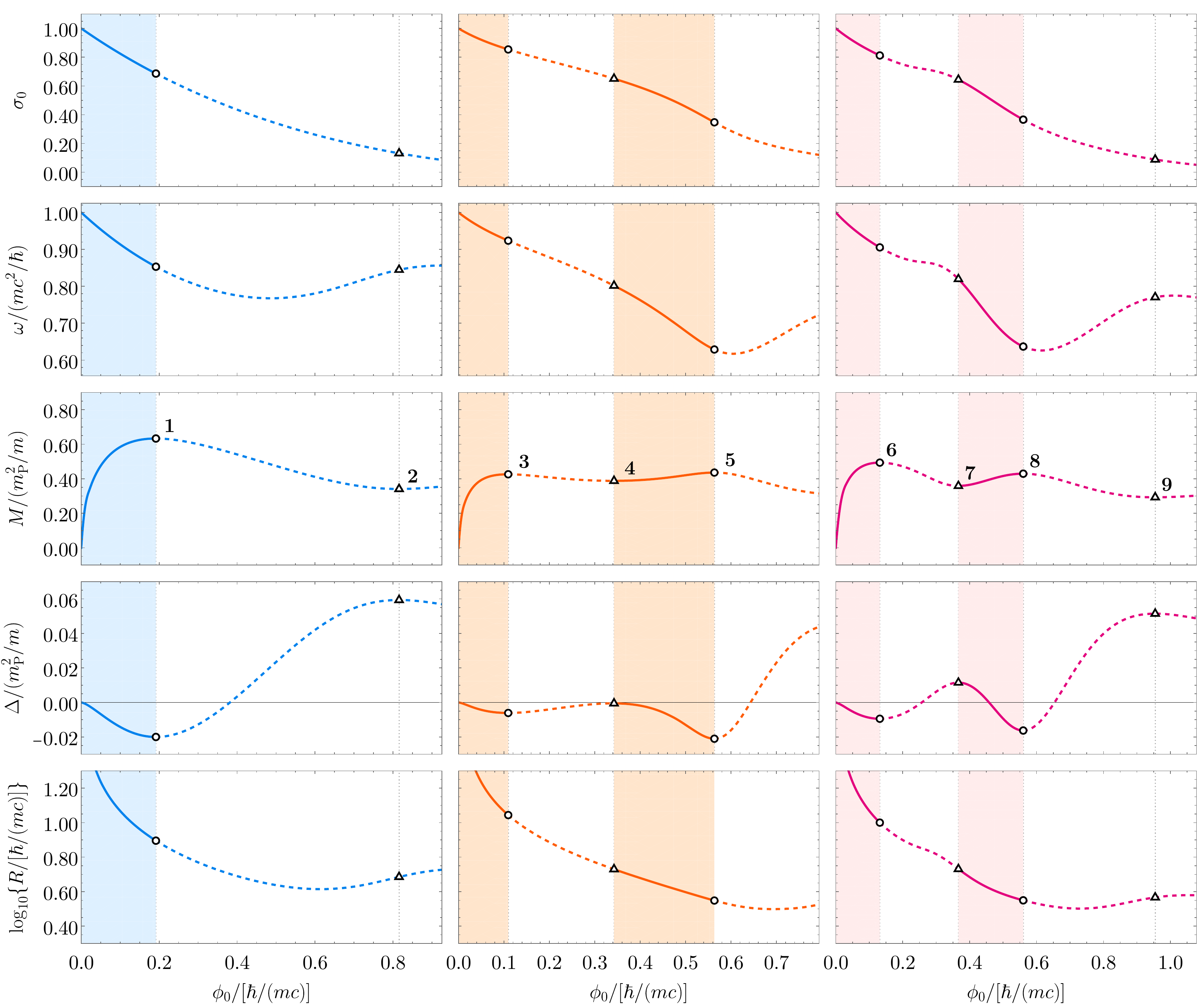}
    \caption{Spherically symmetric spin-0 boson stars in equilibrium. Central value of the metric function $\sigma$, $\sigma_0$ (first row), frequency (second row), mass (third row), binding energy, $\Delta\equiv M-mQ$ (fourth row), and radius (fifth row), all as functions of the shooting parameter $\phi_0$, for mini-boson stars (left column), solitonic boson stars with $v_0=0.20$ (middle column), and axion boson stars with $f_a=0.08$ (right column). The critical points, labeled numerically in the third row (cf.~\autoref{tab:2}), are represented by open circles (maxima) and open triangles (minima). Stars along solid (dashed) curves are stable (unstable), i.e. $\Omega^2>0$ ($\Omega^2<0$) (cf.~\autoref{fig:5.3}).}
    \label{fig:5.1}
\end{figure}

\begin{figure}[ht!]
    \centering
    \makebox[\textwidth]
    {
    \begin{minipage}[]{.32\textwidth}
        \centering
        \includegraphics[scale=0.28]{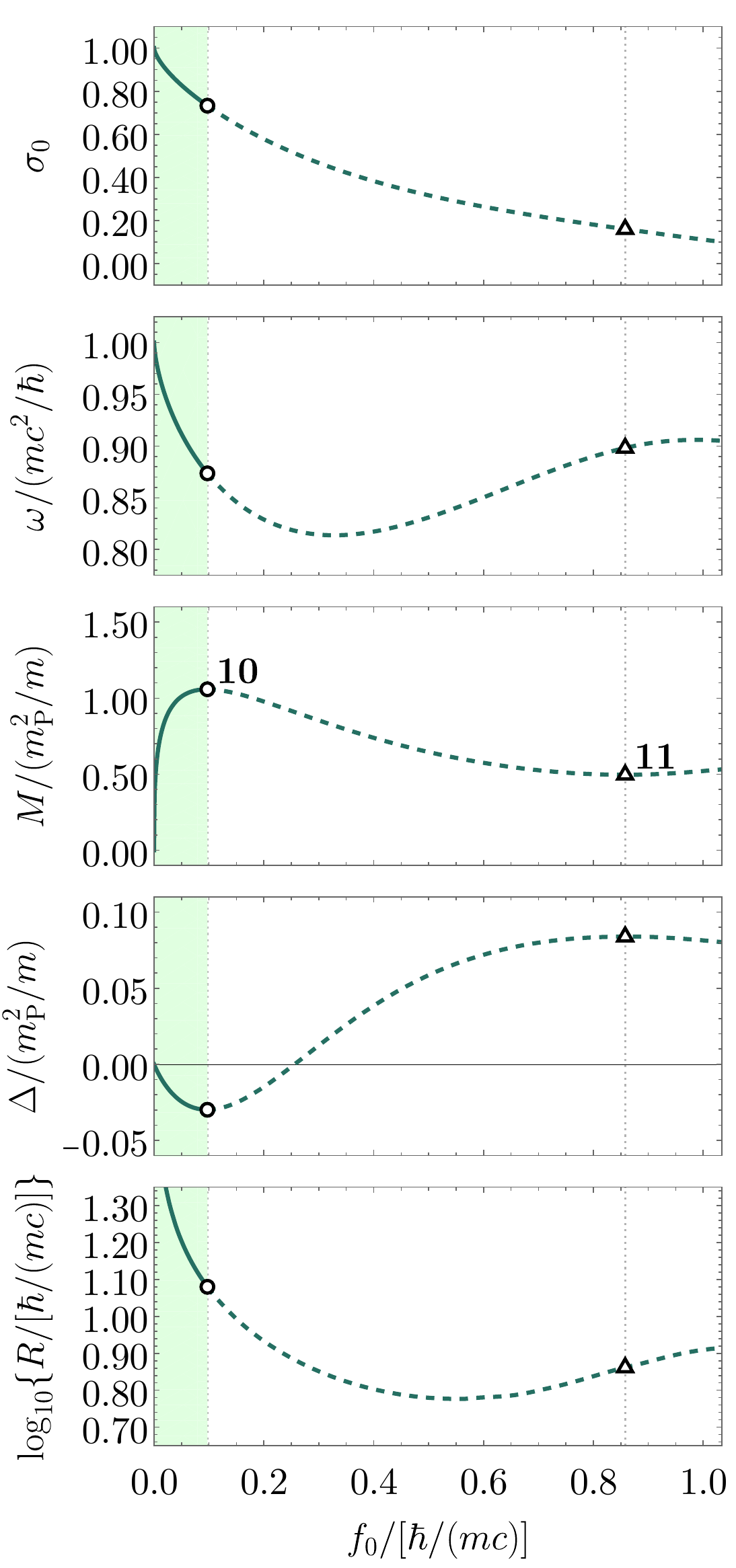}
        \caption{Same as in~\autoref{fig:5.1} but for spherically symmetric spin-1 boson stars in equilibrium (cf.~\autoref{fig:5.4}). $f_0$ plays the role of $\phi_0$ therein.}
        \label{fig:5.2}
    \end{minipage}%
    \quad
    \begin{minipage}[h]{.67\textwidth}
        \vspace{0pt}
        \centering
        \includegraphics[scale=0.28]{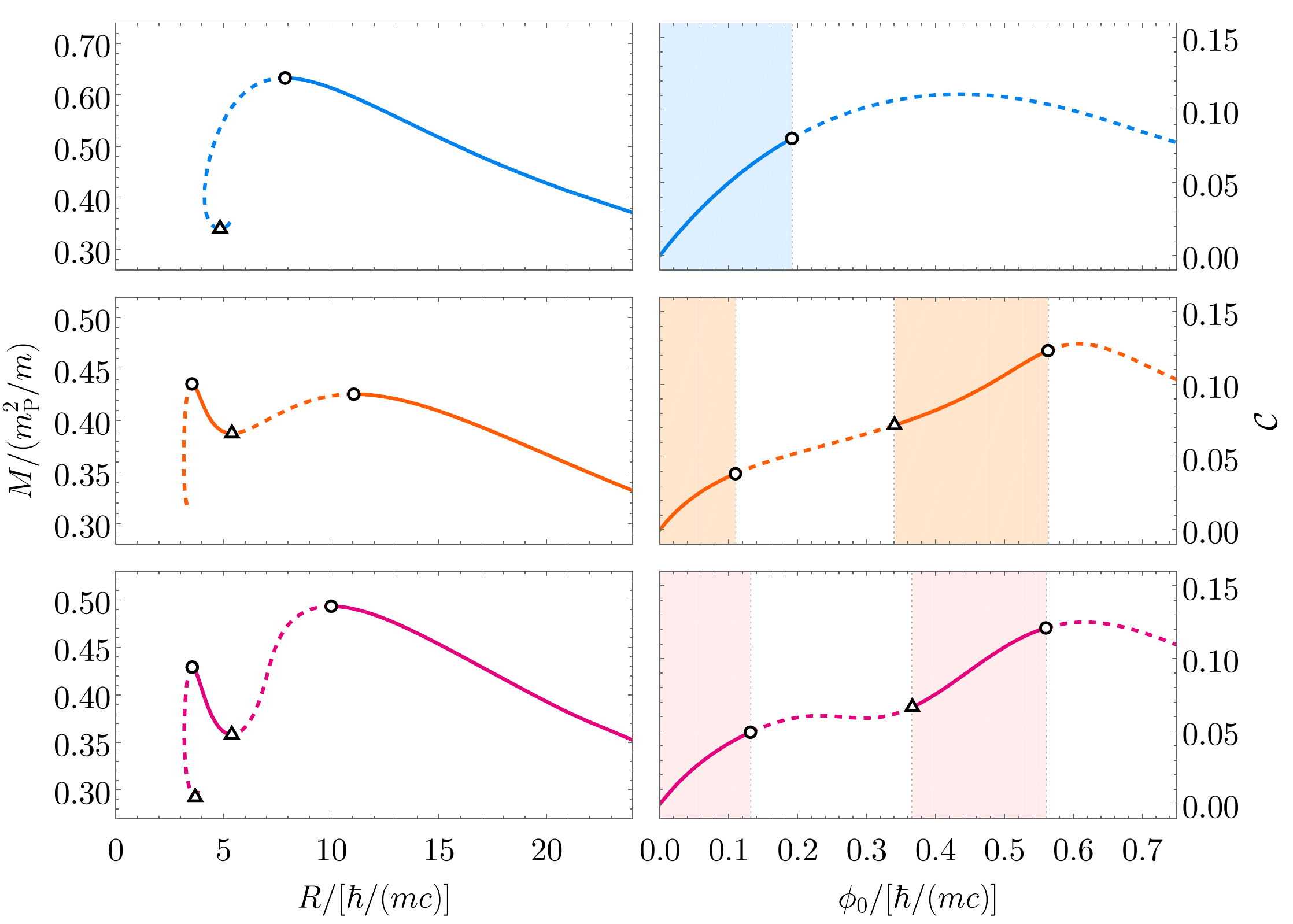}\\
        \vspace{0.5em}
        \includegraphics[scale=0.28]{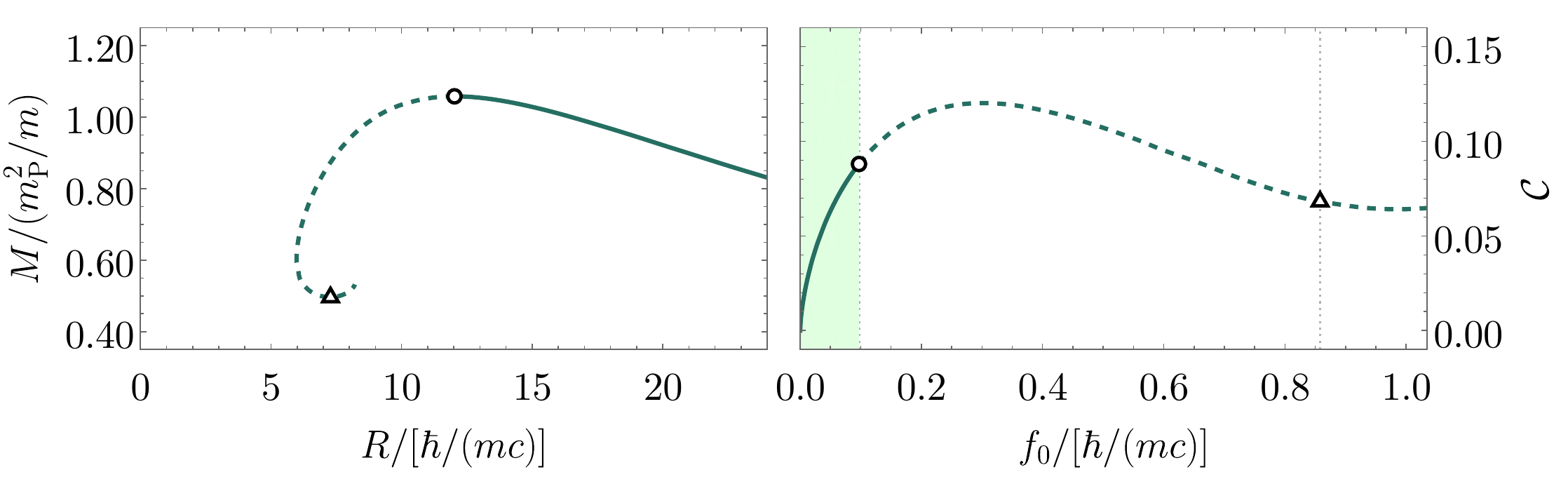}
        \caption{Mass-radius diagram (left column) and compactness as a function of the shooting parameter $\psi_0$ (right column) for the spin-0 (top rows) and spin-1 (bottom row) boson stars in~\autoref{fig:5.1} and~\autoref{fig:5.2}, respectively, where the colors match those therein.}
        \label{fig:5.2b}
    \end{minipage}
    }
\end{figure}

\autoref{fig:5.1} (left column) shows the parameter space of fundamental MBSs (see also~\autoref{fig:5.2b}). Their mass-radius relation is akin to that of neutron stars~\cite{Shapiro:1983du}. The minimum radius of MBSs is $4.12/\mu$. Like neutron stars, they also have a maximum mass, $M_\text{MBSs}^{(\text{max})}\approx 0.633\,m_\text{P}^2/m$. The maximum-mass solution also maximizes the Noether charge, but its binding energy is negative. It is long known that this configuration marks a change in the stability of MBSs. This will be addressed in the following section. 

The general features outlined above might change when adding self-interactions to the scalar-field potential. Recall, however, that both the solitonic and axionic potentials reduce to that of MBSs when $v_0\gg|\Phi|$ and $f_a\gg\sqrt{\hbar}|\Phi|$, respectively. Of particular interest here are cases for which the mass has several critical points with respect to $\phi_0$. This can occur in the complementary regimes $v_0\lesssim|\Phi|$ and $f_a\lesssim\sqrt{\hbar}|\Phi|$, as it is clear from \autoref{fig:5.1} (middle and right columns) for fundamental SBSs with $v_0=0.20$ and fundamental ABSs with $f_a=0.08$, respectively. Their parameter space differs significantly from that of MBSs, displaying new branches, and hence new mass extrema. The inclusion of self-interactions can in particular make the first maximum a local (rather than a global) extremum. Although this is only shown for SBSs, the same is possible for ABSs with sufficiently small values of $f_a$~\cite{Guerra:2019srj}. In the neighborhood of the new maxima lie configurations with negative binding energy. Additionally, self-interacting BSs can be ``smaller" and more compact than MBSs, as shown in~\autoref{fig:5.2b}.

The parameter space of fundamental MBSs and PSs, on the other hand, look very much alike, with $f_0$ playing the role of $\phi_0$ in the latter -- see \autoref{fig:5.2} and \autoref{fig:5.2b} (bottom row). The maximum mass of PSs is significantly larger than that of MBSs, $M_\text{PSs}^{(\text{max})}\approx 1.058\,m_\text{P}^2/m$. As before, PSs can become heavier and more compact when (repulsive) self-interactions are taken into account~\cite{Aoki:2022mdn}, but this case is not considered here.

\subsection{Perturbed solutions}\label{sec:5.2}

Each equilibrium solution is expected to have an infinite countable number of squared perturbation frequencies $\{\Omega^2_n\}_{n=0}^\infty$, labeled by the node number $n$. The lowest-frequency mode is called the fundamental mode, whereas modes with $n>0$ are excited modes. When, for a given solution, the fundamental mode is stable ($\Omega^2_0>0$), then all excited modes are stable, because $\Omega_0^2<\Omega_1^2<\Omega_2^2<\ldots$. When it is unstable, the first excited mode can be either stable or unstable, because $\Omega^2_0<\Omega^2_1$. If stable, all higher-$n$ modes are also stable. If unstable, the second excited mode can be either stable or unstable, because $\Omega^2_1<\Omega^2_2$. The very same reasoning can be generalized to the $n$-th mode. The fundamental mode is thus the most relevant to linear stability. Still, in the following, both fundamental ($n=0$) and first ($n=1$) excited modes are considered. The latter are only used to benchmark the results against those available in the literature (see~\autoref{app:A}).  

\begin{figure}[ht!]
    \centering
    \includegraphics[scale=0.28]{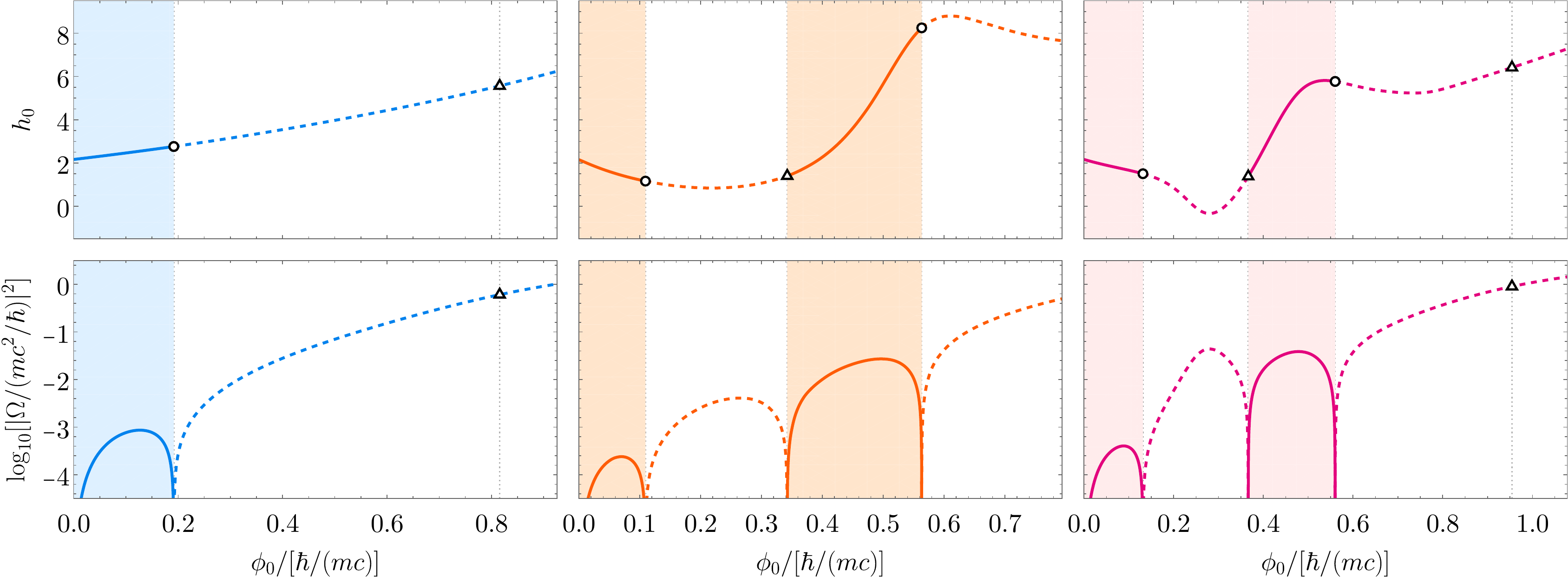}
    \caption{Perturbed spherically symmetric spin-0 boson stars. Central value of the metric function $H_0$, $h_0$ (first row), and squared fundamental frequency $\Omega_0^2$, both as functions of the shooting parameter $\phi_0$, for mini-boson stars (left column), solitonic boson stars with $v_0=0.20$ (middle column), and axion boson stars with $f_a=0.08$ (right column), in a similar fashion to that in~\autoref{fig:5.1}.}
    \label{fig:5.3}
\end{figure}

\begin{figure}[ht!]
    \centering
    \includegraphics[scale=0.28]{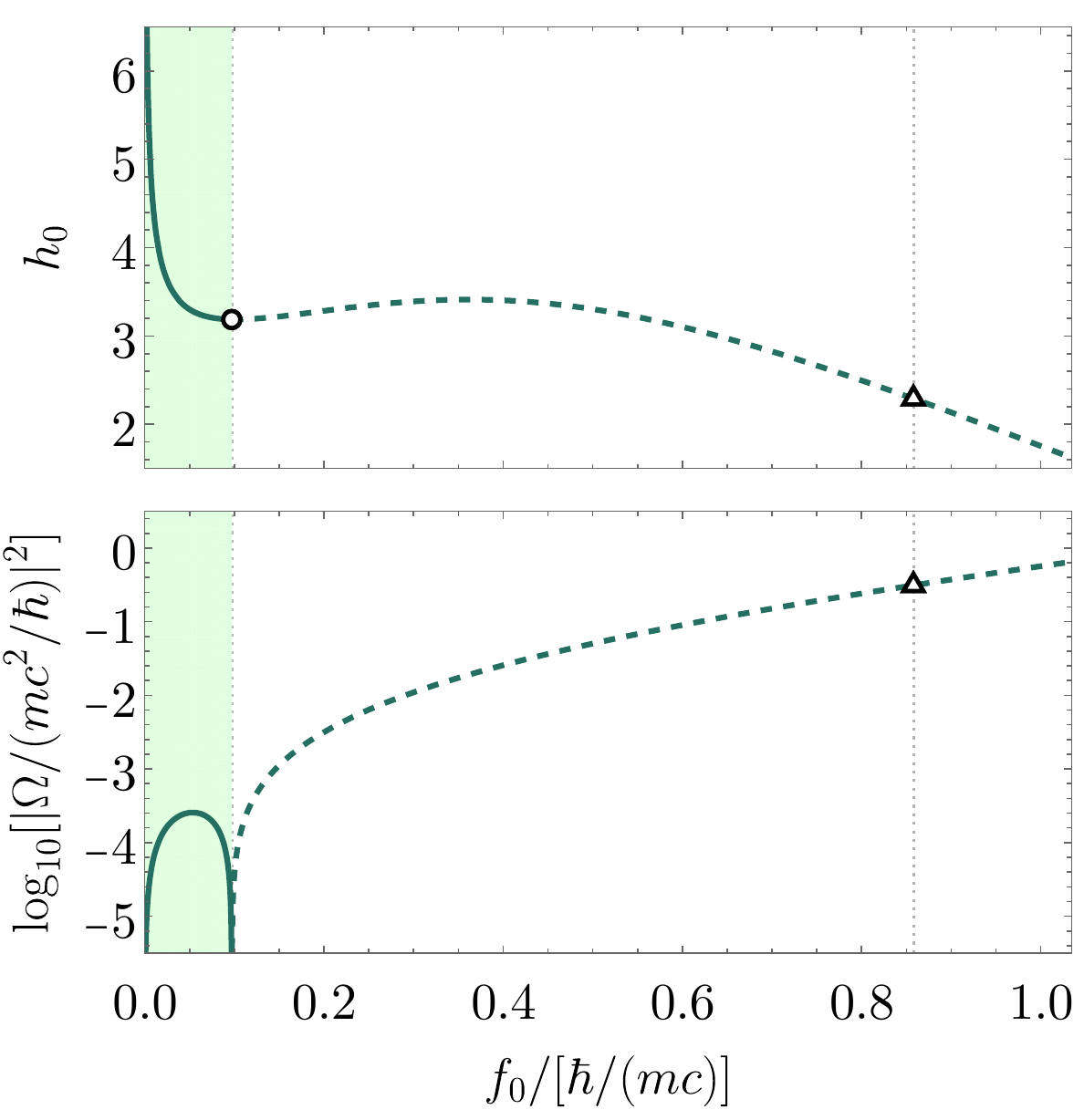}
    \caption{Same as in~\autoref{fig:5.3} but for spherically symmetric spin-1 boson stars. $f_0$ plays the role of $\phi_0$ therein.}
    \label{fig:5.4}
\end{figure}

To find perturbed BSs, one takes an equilibrium solution, provides an initial guess for  $\{h_0,\Omega^2\}$, and solves the boundary value problem defined by Eqs.~\eqref{eq:4.4} and Eqs.~\eqref{eq:4.6} or \eqref{eq:4.7}. The squared fundamental frequencies $\Omega_0^2$ of the equilibrium solutions presented in the previous section are shown in~\autoref{fig:5.3} and ~\autoref{fig:5.4}. As expected, they vanish in the Newtonian limit ($\omega\rightarrow\mu$). For MBSs and PSs, $\Omega^2_0$ first increases with increasing $\phi_0$ and $f_0$, respectively, reaches a maximum value and then drops to zero at the maximum-mass solution. The frequency $\Omega_0$ ranges roughly between $0$ and $10^{-2}$ (in units of $mc^2/\hbar$) in the first branch of both MBSs and PSs, but it is typically larger for the latter. The excited states are also stable. The zero mode ($\Omega^2_0=0$) marks the onset of the instability. Indeed, $\Omega^2_0$ becomes negative as $\phi_0$ and $f_0$ increase beyond their maximum-mass values, with shorter and shorter instability timescales. In summary, both MBSs and PSs are linearly stable against radial perturbations in the first branch but succumb to instabilities otherwise. The dynamical evolution of perturbed MBSs \cite{Seidel:1990jh,Hawley:2000dt} corroborates this result: configurations in the first branch are set into oscillation, dominated by the fundamental frequency $\Omega_0$, which results in a leak of scalar radiation and a decrease in the mass; configurations in the remaining branches either migrate towards the first branch, collapse into BHs or disperse altogether. It is worth noting that the aforementioned dynamical evolutions were performed by enforcing spherical symmetry, which prevents MBSs from decaying into non-spherical configurations in case such decay channels would be dynamically favored. The authors are not aware of any study of non-spherical perturbations of spherically symmetric bosonic stars. Lifting this restriction would make it possible to trigger non-monopolar modes that could change the dynamics. Given that spherical configurations are thought to be the ground state of MBSs, it is likely, though, that their stability properties remain unaltered. As for spherical PSs, recent dynamical evolutions have shown that, in the absence of spatial symmetries or even imposing a $\mathbb{Z}_2$-symmetry, configurations in the first branch, supposed to be stable according to linear perturbation theory, grow a dipolar mode and become prolate PSs~\cite{Herdeiro:2023wqf}. This is because the former are not the true ground states of PSs, as first realized in the Newtonian limit~\cite{Wang:2023tly}.

The parameter space of both MBSs and PSs feature two critical points: a (global) maximum ($\mathbf{1}$ and $\mathbf{10}$, respectively) and a (local) minimum ($\mathbf{2}$ and $\mathbf{11}$, respectively). While the maxima do correspond to a change in the stability of the fundamental mode, it is clear from ~\autoref{fig:5.3} and ~\autoref{fig:5.4} that the minima do not signal any instability-stability transition, but rather represent unstable configurations. The very same conclusions can be drawn by applying the critical point method, i.e. computing the sign of $\text{d}R/\text{d}\psi_0$ at each critical point, as illustrated in~\autoref{tab:2}. These results show that critical points are \textit{not always} zero modes, pinpointing stability reversals (either from stability to instability or vice versa). 

Turning to self-interacting BSs, the spectrum of fundamental frequencies is richer, with (at least) two stable and two unstable branches. As before, the first branch, which contains the Newtonian limit, is stable, whereas the second branch is unstable. The instability timescale has a minimum value, though, which depends on the parameters $\sigma_0$ and $f_a$. The local minima $\mathbf{4}$ and $\mathbf{7}$ correspond to zero modes, meaning that the third branch is also stable. The frequency $\Omega_0$ also attains a maximum value in this branch, about one order of magnitude larger than that of the first branch. The third and fourth branches are separated by the maxima $\mathbf{5}$ and $\mathbf{8}$, for which $\Omega^2_0=0$. The parameter space of ABSs has another critical point, the local minimum $\mathbf{9}$, which is not a zero mode: the fourth and the fifth branches are both unstable. Once again, the critical point method is in agreement with these conclusions, as shown in~\autoref{tab:2}.

\begin{table}[ht!]
    \centering
    \begin{tabular}{cccccccc}
        \toprule
        Critical point & $\dfrac{\psi_0}{\hbar/(mc)}$ & $\dfrac{\omega}{mc^2/\hbar}$ & $\dfrac{M}{m_\text{P}^2/m}$ & $\dfrac{Q}{m_\text{P}^2/m^2}$ & $\dfrac{R}{\hbar/(mc)}$ & $\dfrac{\text{d}R}{\text{d}\psi_0}$ & $\left(\dfrac{\Omega_0}{m c^2/\hbar}\right)^2$ \\
        \midrule
        $\mathbf{1}$ & $0.192$ & $0.853$ & $0.633$ & $0.653$ & $7.86$ & $<0$ & $0$\\
        $\mathbf{2}$ & $0.816$ & $0.845$ & $0.341$ & $0.281$ & $4.84$ & $>0$ & $-0.606$\\
        \midrule
        $\mathbf{3}$ & $0.109$ & $0.924$ & $0.426$ & $0.432$ & $11.1$ & $<0$ & $0$\\
        $\mathbf{4}$ & $0.342$ & $0.802$ & $0.388$ & $0.388$ & $5.37$ & $<0$ & $0$\\
        $\mathbf{5}$ & $0.563$ & $0.629$ & $0.436$ & $0.457$ & $3.54$ & $<0$ & $0$\\
        \midrule
        $\mathbf{6}$ & $0.132$ & $0.905$ & $0.493$ & $0.503$ & $10.0$ & $<0$ & $0$\\
        $\mathbf{7}$ & $0.366$ & $0.820$ & $0.358$ & $0.347$ & $5.39$ & $<0$ & $0$\\
        $\mathbf{8}$ & $0.560$ & $0.637$ & $0.429$ & $0.445$ & $3.55$ & $<0$ & $0$\\
        $\mathbf{9}$ & $0.955$ & $0.770$ & $0.294$ & $0.241$ & $3.69$ & $>0$ & $-0.900$ \\
        \midrule
        $\mathbf{10}$ & $0.0971$ & $0.874$ & $1.06$ & $1.09$ & $12.0$ & $<0$ & $0$\\
        $\mathbf{11}$ & $0.858$ & $0.898$ & $0.496$ & $0.412$ & $7.28$ & $>0$ & $-0.312$\\
        \bottomrule
    \end{tabular}
    \caption{Critical points identified in~\autoref{fig:5.1} and \autoref{fig:5.2}.}
    \label{tab:2}
\end{table}

\section{Conclusion}\label{sec:6}

Dynamical stability is a central notion in physics. A system in dynamic equilibrium is said to be dynamically stable if, when perturbed, all physical quantities associated with the perturbation remain bounded in time, and, ultimately, dynamic equilibrium is restored. Otherwise, the system is said to be dynamically unstable. This notion encompasses those of linear dynamical stability as well as mode stability. Although both refer to linear perturbations (i.e. sufficiently small disturbances), the latter concerns mode (i.e. fixed-frequency) perturbations. Mode stability guarantees that the system does not have exponentially growing mode solutions (with $\text{Im}\,\Omega>0$).

Dynamical stability implies linear dynamical stability, which in turn implies mode stability. It is then natural to assess the dynamical stability of a system by first studying its mode stability. This amounts to considering mode solutions to the equations governing linear perturbations. There has been a major effort to establish the mode stability of several compact objects, such as fluid stars, BSs, and BHs. This problem is generally intractable unless symmetries are imposed on the system. Spherical symmetry is frequently the starting point for such analysis, as it greatly simplifies the perturbation equations. A further simplification is to consider monopolar perturbations, i.e. study radial stability. Such oversimplification -- narrowing dynamical stability down to radial stability -- may seem pointless. However, as far as spherically symmetric systems are concerned, radial stability analysis can provide a good proxy for linear dynamics, as reported in \cite{Hawley:2000dt,Kain:2021rmk} for BSs. 

This paper addressed the radial stability of spherically symmetric mini-, solitonic, and axionic BSs as well as PSs. It focused on the numerical computation of their fundamental squared perturbation frequency. Previous works mostly looked into MBSs, and results for SBSs and ABSs were sparse or even lacking in the literature.\footnote{To the best of the author's knowledge, the radial stability of SBSs was only partially addressed in~\cite{Ildefonso:2023qty}.} The addition of self-interactions to the scalar-field potential turns the parameter space more involved, with more branches and also zero-frequency modes. As for PSs, their radial stability was preliminarily studied in~\cite{Brito:2015pxa}, wherein the fundamental mode frequency was presented for configurations in the neighborhood of the maximum-mass PS. These results were extended herein to the whole parameter space of PSs. Their spectrum turns out to bear a close resemblance to that of MBSs, featuring a stable branch and an unstable branch, divided by the maximum-mass solution. Nonetheless, spherically symmetric PSs along the stable branch (with respect to spherical perturbations) are not dynamically stable, as they decay into prolate configurations, the true ground states of static PSs~\cite{Herdeiro:2023wqf}. This is a remarkable example of how (radial) mode stability may not translate into dynamical stability. In light of this result, it would be of particular interest to study dipolar perturbations of prolate PSs. 

\begin{figure}[ht!]
    \centering
    \includegraphics[scale=0.30]{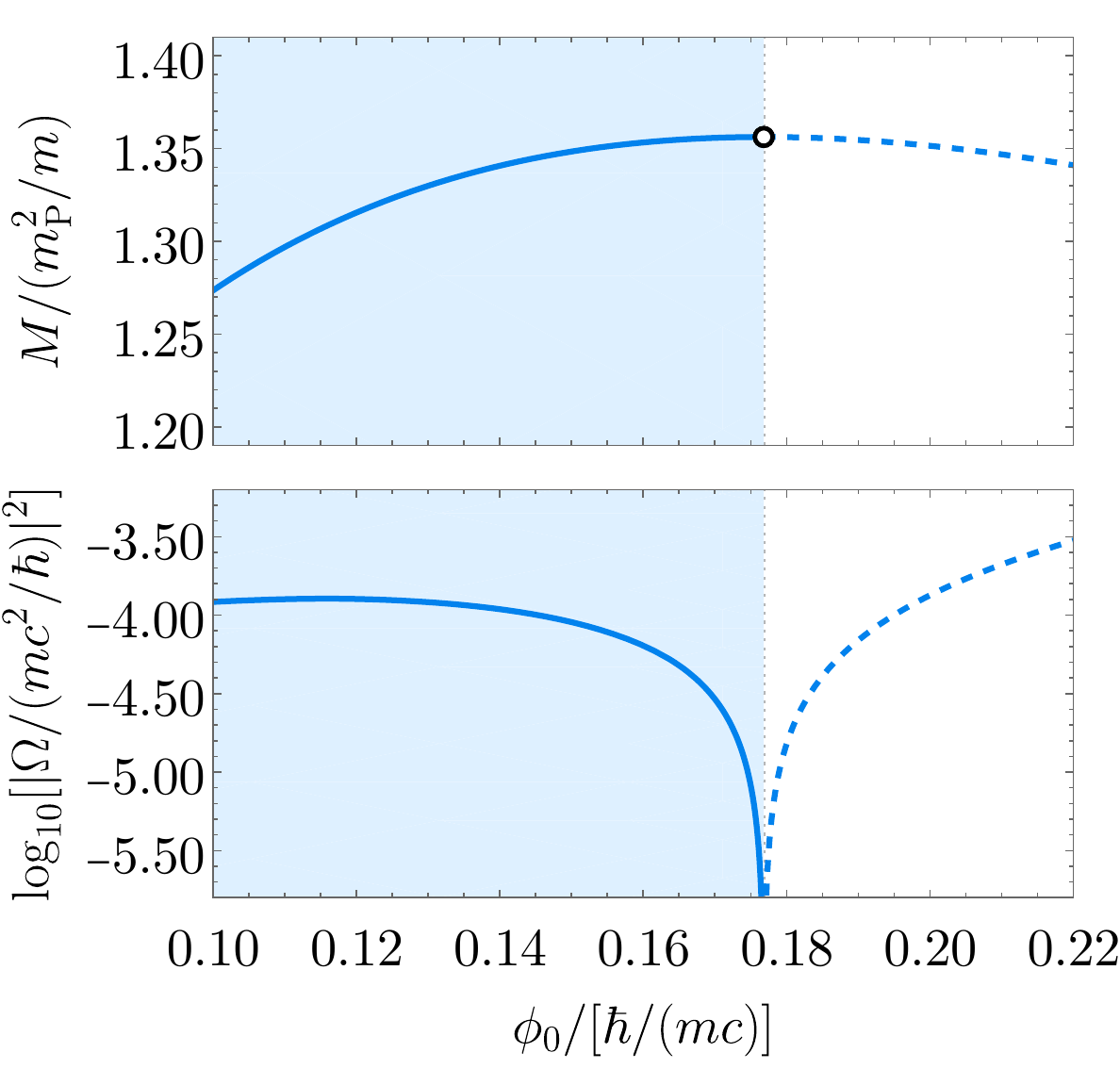}
    \caption{One-node spherically symmetric mini-boson stars. Mass (top row) and squared fundamental frequency $\Omega_0^2$ (bottom row), both as functions of the shooting parameter $\phi_0$.}
    \label{fig:6.1}
\end{figure}

Another example of the aforementioned contrast is that of radially excited MBSs, i.e. configurations with at least one node in the radial direction. \autoref{fig:6.1} shows the parameter space of one-node MBSs is akin to that of the ground state, with a stable branch connecting the Newtonian limit to the maximum-mass solution. Such putatively stable configurations turn out to be \textit{dynamically unstable}, either decaying to the ground state or collapsing into a BH~\cite{Balakrishna:1997ej}. In other words, mode stability may not persist when non-linear effects are taken into account.

The main take-home message from this study is that $\text{d}M/\text{d}\psi_0=0$ is not a sufficient condition for the existence of a zero-frequency mode and, therefore, for a stability-instability transition (as far as spherically symmetric bosonic stars are concerned\footnote{See \cite{Takami:2011zc,Weih:2017mcw} for a similar discussion on the critical point method applied to rotating relativistic stars.}). This statement is not always made explicit in the literature -- rather the opposite.\footnote{Exceptions exist, though. This subtlety has been properly addressed in the original work on ABSs~\cite{Guerra:2019srj}.} As a result, the condition $\text{d}M/\text{d}\psi_0=0$ can be easily misread as equivalent to $\Omega^2=0$. \autoref{tab:2} makes the difference clear: all solutions listed therein satisfy $\text{d}M/\text{d}\psi_0=0$ but only those with $\Omega_0^2=0$ pinpoint changes in stability across the corresponding parameter space, as shown in \autoref{fig:5.3} and \autoref{fig:5.4}. The authors hope the examples presented herein help clear up any misconceptions about the meaning of critical points on a plot of $M$ vs. $\psi_0$, as they can lead to fallacious conclusions. 

\section*{Acknowledgements}

The authors would like to thank Richard Brito for discussions on the perturbation theory of PSs, and Eugen Radu for constructive comments and feedback on the manuscript. C. L. B. thanks the University of Aveiro for the kind hospitality during the development of part of this work. 

This work is supported by the Center for Research and Development in Mathematics and Applications (CIDMA) through the Portuguese Foundation for Science and Technology (FCT -- Fundaç\~ao para a Ci\^encia e a Tecnologia) through projects: UIDB/04106/2020 (DOI identifier \url{https://doi.org/10.54499/UIDB/04106/2020}); UIDP/04106/2020 (DOI identifier \url{https://doi.org/10.54499/UIDP/04106/2020});  PTDC/FIS-AST/3041/2020 (DOI identifier \url{http://doi.org/10.54499/PTDC/FIS-AST/3041/2020}); CERN/FIS-PAR/0024/2021 (DOI identifier \url{http://doi.org/10.54499/CERN/FIS-PAR/0024/2021}); and 2022.04560.PTDC (DOI identifier \url{https://doi.org/10.54499/2022.04560.PTDC}). This work has further been supported by the European Horizon Europe staff exchange (SE) programme HORIZON-MSCA-2021-SE-01 Grant No.\ NewFunFiCO-101086251. N. M. S. is supported by the FCT grant SFRH/BD/143407/2019 (DOI identifier \url{https://doi.org/10.54499/SFRH/BD/143407/2019}). C. L. B. acknowledges Conselho Nacional de Desenvolvimento Cient\'ifico e Tecnol\'ogico (CNPq) and Funda\c{c}\~ao Amaz\^onia Paraense de Amparo \`a Pesquisa (FAPESPA), from Brazil, for partial financial support.

\appendix


\section{Comparison with the literature}
\label{app:A}

\begin{figure}[ht!]
    \centering
    \includegraphics[scale=0.28]{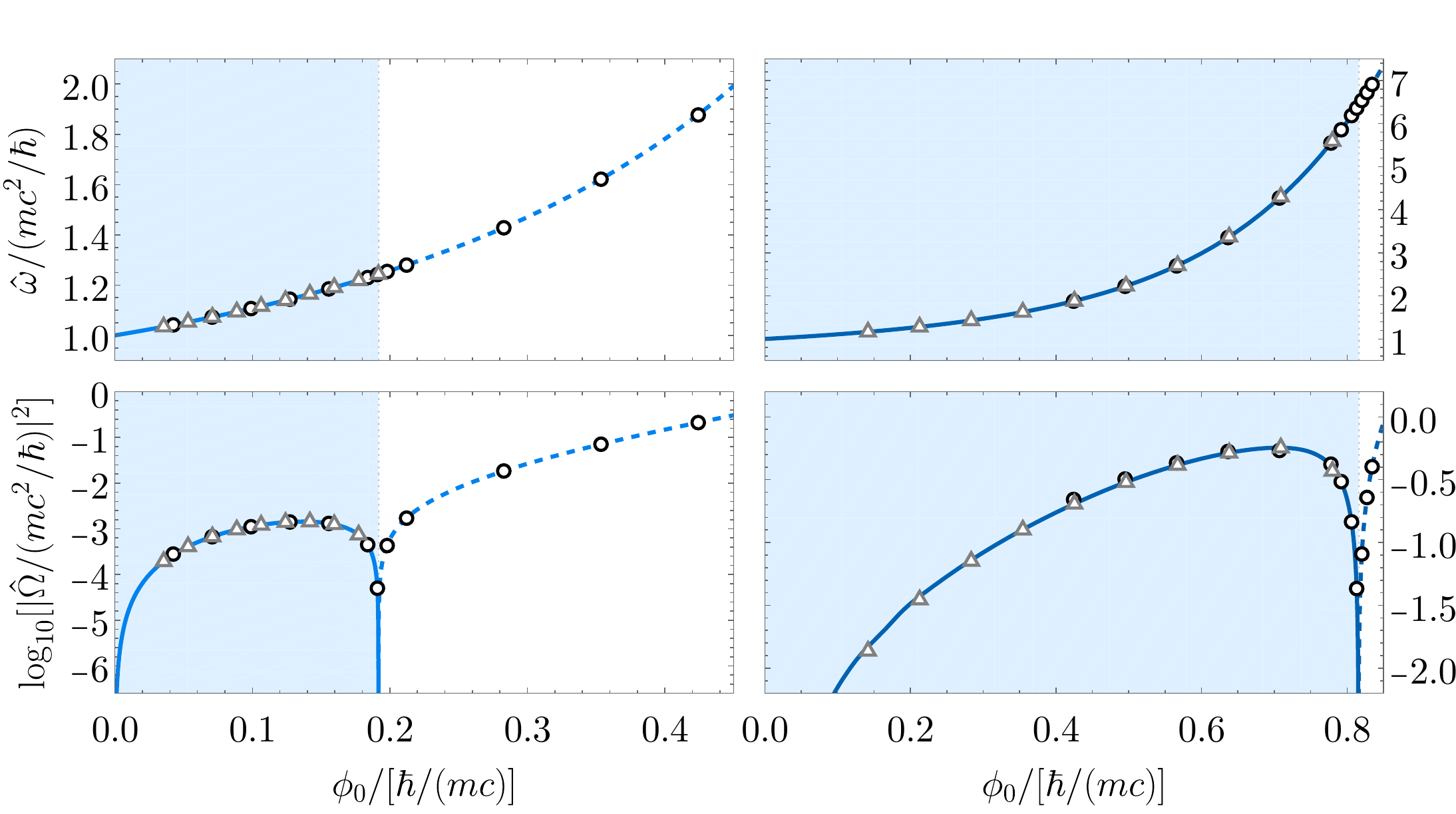}
    \caption{Spherically symmetric mini-boson stars. Rescaled frequencies $\{\hat{\omega},\hat{\Omega}\}$ of the fundamental (left column) and first excited (right column) modes, both as a function of the shooting parameter $\phi_0$. The black open circles and the gray open triangles are the data reported in~\cite{Hawley:2000dt} and~\cite{Kain:2021rmk}, respectively.}
    \label{fig:8}
\end{figure}

For completeness, some of the results presented in~\autoref{sec:5} are benchmarked against those available in the literature for the fundamental and first excited modes of MBSs~\cite{Hawley:2000dt,Kain:2021rmk}. For that purpose, it is convenient to define the rescaled frequencies
    \begin{align}
        \hat{\omega}\equiv\frac{\omega}{\sigma_0}\ ,
        \qquad
        \hat{\Omega}\equiv\frac{\Omega}{\sigma_0}\ ,
    \end{align}
as they can be readily compared to those in~\cite{Hawley:2000dt}. \Cref{tab:3,tab:4} (\Cref{tab:5,tab:6}) list relevant quantities of this work's solutions with values of $\phi_0$ matching those selected in~\cite{Hawley:2000dt} (\cite{Kain:2021rmk}) for the fundamental and first excited state, respectively. The last two columns of each table contains the values of $\hat{\omega}$ and $\hat{\Omega}_0^2$ ($\omega$ and $\Omega_0^2$) reported therein. In general, the agreement is excellent, as clearly shown in~\autoref{fig:8}, although there are some discrepancies in $\Omega_0^2$ or $\hat{\Omega}_0^2$. This difference is likely to be linked to the cumulative errors when computing the perturbed solutions. This computation amounts to solving the ``pulsation equations" using the equilibrium solutions, which, having been found numerically, already contain errors. 

\begin{landscape}
\begin{table}
    \centering
    \begin{tabular}{ccccccccc||cc}
        \toprule
        $\dfrac{\phi_0}{\sqrt{2}\hbar/(mc)}$ & $\sigma_0$ & $\dfrac{\hat{\omega}}{mc^2/\hbar}$ & $\dfrac{\omega}{mc^2/\hbar}$ & $\dfrac{M}{m_\text{P}^2/m}$ & $\dfrac{Q}{m_\text{P}^2/m^2}$ & $h_0$ & $\left(\dfrac{\Omega_0}{m c^2/\hbar}\right)^2$ & $\left(\dfrac{\hat{\Omega}_0}{m c^2/\hbar}\right)^2$ & $\left(\dfrac{\hat{\omega}}{m c^2/\hbar}\right)^2$~\cite{Hawley:2000dt} & $\left(\dfrac{\hat{\Omega}_0}{m c^2/\hbar}\right)^2$~\cite{Hawley:2000dt}\\
        \midrule
        0.06 & 0.9222 & 1.0417 & 0.9606 & 0.4468 & 0.4523 & 0.1821 & $2.310\times10^{-4}$ & $2.716\times10^{-4}$ & 1.0417 & $2.8\times10^{-4}$\\
        0.10 & 0.8732 & 1.0727 & 0.9367 & 0.5326 & 0.5427 & 0.1889 & $5.133\times10^{-4}$ & $6.732\times10^{-4}$ & 1.0727 & $6.7\times10^{-4}$\\
        0.14 & 0.8263 & 1.1067 & 0.9144 & 0.5827 & 0.5968 & 0.1958 & $7.597\times10^{-4}$ & $1.113\times10^{-3}$ & 1.1067 & $1.11\times10^{-3}$\\
        0.18 & 0.7814 & 1.1439 & 0.8939 & 0.6118 & 0.6289 & 0.2029 & $8.629\times10^{-4}$ & $1.413\times10^{-3}$ & 1.1440 & $1.41\times10^{-3}$\\
        0.22 & 0.7384 & 1.1849 & 0.8749 & 0.6271 & 0.6462 & 0.2102 & $7.162\times10^{-4}$ & $1.314\times10^{-3}$ & 1.1849 & $1.31\times10^{-3}$\\
        0.26 & 0.6973 & 1.2299 & 0.8576 & 0.6328 & 0.6527 & 0.2177 & $2.147\times10^{-4}$ & $4.415\times10^{-4}$ & 1.2299 & $4.5\times10^{-4}$\\
        0.27 & 0.6872 & 1.2419 & 0.8535 & 0.6330 & 0.6530 & 0.2196 & $2.228\times10^{-5}$ & $4.717\times10^{-5}$ & 1.2419 & $0.5\times10^{-4}$\\
        0.28 & 0.6773 & 1.2541 & 0.8495 & 0.6328 & 0.6528 & 0.2215 & $-2.010\times10^{-4}$ & $-4.380\times10^{-4}$ & 1.2542 & $-0.43\times10^{-3}$\\
        0.30 & 0.6578 & 1.2796 & 0.8418 & 0.6315 & 0.6512 & 0.2254 & $-7.456\times10^{-4}$ & $-1.723\times10^{-3}$ & 1.2796 & $-1.71\times10^{-3}$\\
        0.40 & 0.5664 & 1.4281 & 0.8088 & 0.6088 & 0.6236 & 0.2456 & $-5.896\times10^{-3}$ & $-1.838\times10^{-2}$ & 1.4281 & $-1.84\times10^{-2}$\\
        0.50 & 0.4843 & 1.6215 & 0.7853 & 0.5703 & 0.5752 & 0.2672 & $-1.622\times10^{-2}$ & $-7.087\times10^{-2}$ & 1.6215 & $-7.09\times10^{-2}$\\
        0.60 & 0.4109 & 1.8777 & 0.7715 & 0.5248 & 0.5167 & 0.2900 & $-3.568\times10^{-2}$ & $-2.114\times10^{-1}$ & 1.8777 & $-2.11\times10^{-1}$\\
        \bottomrule
    \end{tabular}
    \caption{Comparison between this work's results and data reported in \cite{Hawley:2000dt} for the fundamental mode (last two columns) of mini-boson stars.}
    \label{tab:3}
\end{table}
\end{landscape}

\begin{landscape}
\begin{table}
    \centering
    \begin{tabular}{ccccccccc||cc}
        \toprule
        $\dfrac{\phi_0}{\sqrt{2}\hbar/(mc)}$ & $\sigma_0$ & $\dfrac{\hat{\omega}}{mc^2/\hbar}$ & $\dfrac{\omega}{mc^2/\hbar}$ & $\dfrac{M}{m_\text{P}^2/m}$ & $\dfrac{Q}{m_\text{P}^2/m^2}$ & $h_0$ & $\left(\dfrac{\Omega_0}{m c^2/\hbar}\right)^2$ & $\left(\dfrac{\hat{\Omega}_0}{m c^2/\hbar}\right)^2$ & $\left(\dfrac{\hat{\omega}}{m c^2/\hbar}\right)^2$~\cite{Hawley:2000dt} & $\left(\dfrac{\hat{\Omega}_0}{m c^2/\hbar}\right)^2$~\cite{Hawley:2000dt}\\
        \midrule
        0.60 & 0.4108 & 1.8767 & 0.7715 & 0.5250 & 0.5169 & 0.3640 & $3.409\times10^{-2}$ & $2.019\times10^{-1}$ & 1.8777 & 0.22\\
        0.70 & 0.3453 & 2.2240 & 0.7675 & 0.4773 & 0.4548 & 0.4011 & $3.581\times10^{-2}$ & $3.003\times10^{-1}$ & 2.2230 & 0.32\\
        0.80 & 0.2871 & 2.6960 & 0.7741 & 0.4314 & 0.3953 & 0.4434 & $3.390\times10^{-2}$ & $4.113\times10^{-1}$ & 2.6963 & 0.43\\
        0.90 & 0.2355 & 3.3536 & 0.7899 & 0.3908 & 0.3434 & 0.4915 & $2.860\times10^{-2}$ & $5.155\times10^{-1}$ & 3.3536 & 0.53\\
        1.00 & 0.1902 & 4.2714 & 0.8123 & 0.3598 & 0.3046 & 0.5439 & $2.055\times10^{-2}$ & $5.682\times10^{-1}$ & 4.2714 & 0.54\\
        1.10 & 0.1505 & 5.5470 & 0.8351 & 0.3428 & 0.2839 & 0.5956 & $9.284\times10^{-3}$ & $4.096\times10^{-1}$ & 5.5471 & 0.42\\
        1.12 & 0.1433 & 5.8554 & 0.8390 & 0.3414 & 0.2822 & 0.6054 & $6.222\times10^{-3}$ & $3.031\times10^{-1}$ & 5.8555 & $3.05\times10^{-1}$\\
        1.14 & 0.1362 & 6.1841 & 0.8425 & 0.3407 & 0.2813 & 0.6150 & $2.705\times10^{-3}$ & $1.457\times10^{-1}$ & 6.1842 & $1.46\times10^{-1}$\\
        1.15 & 0.1328 & 6.3566 & 0.8442 & 0.3405 & 0.2812 & 0.6197 & $7.466\times10^{-4}$ & $4.233\times10^{-2}$ & 6.3566 & $4.30\times10^{-2}$\\
        1.16 & 0.1294 & 6.5346 & 0.8457 & 0.3405 & 0.2812 & 0.6243 & $-1.360\times10^{-3}$ & $-8.123\times10^{-2}$ & 6.5346 & $-8.11\times10^{-2}$\\
        1.17 & 0.1261 & 6.7184 & 0.8471 & 0.3407 & 0.2814 & 0.6289 & $-3.627\times10^{-3}$ & $-2.281\times10^{-1}$ & 6.7184 & $-2.28\times10^{-1}$\\
        1.18 & 0.1228 & 6.9082 & 0.8485 & 0.3410 & 0.2818 & 0.6335 & $-6.064\times10^{-3}$ & $-4.020\times10^{-1}$ & 6.9083 & $-4.01\times10^{-1}$\\
        \bottomrule
    \end{tabular}
    \caption{Comparison between this work's results and data reported in \cite{Hawley:2000dt} for the first excited mode (last two columns) of mini-boson stars.}
    \label{tab:4}
\end{table}
\end{landscape}

\begin{landscape}
\begin{table}
    \centering
    \begin{tabular}{cccccccc||cc}
        \toprule
        $\dfrac{\phi_0}{\sqrt{4\pi}\hbar/(mc)}$ & $\sigma_0$ & $\dfrac{\hat{\omega}}{mc^2/\hbar}$ & $\dfrac{\omega}{mc^2/\hbar}$ & $\dfrac{M}{m_\text{P}^2/m}$ & $\dfrac{Q}{m_\text{P}^2/m^2}$ & $h_0$ & $\left(\dfrac{\Omega_0}{m c^2/\hbar}\right)^2$ & $\dfrac{\omega}{m c^2/\hbar}$~\cite{Kain:2021rmk} & $\left(\dfrac{\Omega_0}{m c^2/\hbar}\right)^2$~\cite{Kain:2021rmk}\\
        \midrule
        0.010 & 0.9346 & 1.0344 & 0.9668 & 0.4167 & 0.4211 & 0.1804 & $1.693\times10^{-4}$ & 0.9668 & $1.691\times10^{-4}$\\
        0.015 & 0.9033 & 1.0531 & 0.9513 & 0.4853 & 0.4926 & 0.1846 & $3.345\times10^{-4}$ & 0.9513 & $3.344\times10^{-4}$\\
        0.020 & 0.8729 & 1.0729 & 0.9365 & 0.5331 & 0.5432 & 0.1889 & $5.152\times10^{-4}$ & 0.9365 & $5.140\times10^{-4}$\\
        0.025 & 0.8432 & 1.0939 & 0.9224 & 0.5673 & 0.5800 & 0.1932 & $6.806\times10^{-4}$ & 0.9224 & $6.798\times10^{-4}$\\
        0.030 & 0.8144 & 1.1160 & 0.9089 & 0.5919 & 0.6069 & 0.1976 & $8.046\times10^{-4}$ & 0.9089 & $8.045\times10^{-4}$\\
        0.035 & 0.7864 & 1.1395 & 0.8961 & 0.6093 & 0.6261 & 0.2021 & $8.617\times10^{-4}$ & 0.8961 & $8.617\times10^{-4}$\\
        0.040 & 0.7591 & 1.1645 & 0.8840 & 0.6211 & 0.6394 & 0.2066 & $8.255\times10^{-4}$ & 0.8840 & $8.253\times10^{-4}$\\
        0.045 & 0.7326 & 1.1909 & 0.8724 & 0.6284 & 0.6477 & 0.2112 & $6.698\times10^{-4}$ & 0.8724 & $6.697\times10^{-4}$\\
        0.050 & 0.7067 & 1.2190 & 0.8615 & 0.6321 & 0.6520 & 0.2159 & $3.693\times10^{-4}$ & 0.8615 & $3.693\times10^{-4}$\\
        \bottomrule
    \end{tabular}
    \caption{Comparison between this work's results and data reported in \cite{Kain:2021rmk} for the fundamental mode (last two columns) of mini-boson stars.}
    \label{tab:5}
\end{table}

\begin{table}
    \centering
    \begin{tabular}{cccccccc||cc}
        \toprule
        $\dfrac{\phi_0}{\sqrt{4\pi}\hbar/(mc)}$ & $\sigma_0$ & $\dfrac{\hat{\omega}}{mc^2/\hbar}$ & $\dfrac{\omega}{mc^2/\hbar}$ & $\dfrac{M}{m_\text{P}^2/m}$ & $\dfrac{Q}{m_\text{P}^2/m^2}$ & $h_0$ & $\left(\dfrac{\Omega_0}{m c^2/\hbar}\right)^2$ & $\dfrac{\omega}{m c^2/\hbar}$~\cite{Kain:2021rmk} & $\left(\dfrac{\Omega_0}{m c^2/\hbar}\right)^2$~\cite{Kain:2021rmk}\\
        \midrule
        0.04 & 0.7591 & 1.1645 & 0.8840 & 0.6212 & 0.6394 & 0.2521 & $8.530\times10^{-3}$ & 0.8840 & $7.892\times10^{-3}$\\
        0.06 & 0.6571 & 1.2806 & 0.8415 & 0.6314 & 0.6511 & 0.2762 & $1.609\times10^{-2}$ & 0.8415 & $1.517\times10^{-2}$\\
        0.08 & 0.5655 & 1.4299 & 0.8085 & 0.6084 & 0.6231 & 0.3020 & $2.319\times10^{-2}$ & 0.8085 & $2.277\times10^{-2}$\\
        0.10 & 0.4833 & 1.6234 & 0.7851 & 0.5700 & 0.5748 & 0.3315 & $2.961\times10^{-2}$ & 0.7851 & $2.943\times10^{-2}$\\
        0.12 & 0.4097 & 1.8814 & 0.7713 & 0.5242 & 0.5159 & 0.3646 & $3.414\times10^{-2}$ & 0.7713 & $3.406\times10^{-2}$\\
        0.14 & 0.3442 & 2.2315 & 0.7676 & 0.4764 & 0.4537 & 0.4018 & $3.580\times10^{-2}$ & 0.7677 & $3.570\times10^{-2}$\\
        0.16 & 0.2859 & 2.7078 & 0.7743 & 0.4305 & 0.3941 & 0.4444 & $3.382\times10^{-2}$ & 0.7743 & $3.379\times10^{-2}$\\
        0.18 & 0.2344 & 3.3721 & 0.7904 & 0.3900 & 0.3423 & 0.4928 & $2.844\times10^{-2}$ & 0.7904 & $2.841\times10^{-2}$\\
        0.20 & 0.1891 & 4.3001 & 0.8130 & 0.3592 & 0.3038 & 0.5453 & $2.030\times10^{-2}$ & 0.8130 & $2.028\times10^{-2}$\\
        0.22 & 0.1495 & 5.5908 & 0.8357 & 0.3426 & 0.2836 & 0.5970 & $8.862\times10^{-3}$ & 0.8357 & $8.862\times10^{-3}$\\
        \bottomrule
    \end{tabular}
    \caption{Comparison between this work's results and data reported in \cite{Kain:2021rmk} for the first excited mode (last two columns) of mini-boson stars.}
    \label{tab:6}
\end{table}
\end{landscape}

\printbibliography

\end{document}